# Achieving electrode smoothing by controlling the nucleation phase of metal deposition through polymer-substrate binding

Ying Xia, Duo Song, Mingyi Zhang, Zheming Wang, Chenyang Shi, Jingshan S. Du, Sun Hae Ra Shin, Mark H. Engelhard, Praveen K. Thallapally, Christine A. Orme, Jinhui Tao, Maria L. Sushko,* James. J. De Yoreo,* and Jun Liu*

Y. Xia, J. J. De Yoreo, J. Liu
Department of Materials Science and Engineering
University of Washington
Seattle, WA 98115, USA
E-mail: jliuuw1@uw.edu

Y. Xia, D. Song, M. Zhang, Z. Wang, C. Shi, J. S. Du, J. Tao, M. L. Sushko, J. J. De Yoreo
Physical and Computational Sciences Directorate
Pacific Northwest National Laboratory
Richland, WA 99354, USA
E-mail: James.DeYoreo@pnnl.gov
E-mail: Maria.Sushko@pnnl.gov

S. H. R. Shin, M. H. Engelhard, P. K. Thallapally, J. Liu
Energy and Environment Directorate
Pacific Northwest National Laboratory
Richland, WA 99354, USA

C. A. Orme
Physical and Life Sciences Directorate
Lawrence Livermore National Laboratory
7000 East Avenue, Livermore, CA 94550, USA

**Abstract:**

Polymer additives [like polyethylene oxide (PEO)] are widely used for smooth electrode deposition in aqueous zinc and a number of other battery systems currently investigated for energy storage applications. However, the precise mechanism by which they regulate morphology and suppress dendrite formation remains unclear. In this study, we address this knowledge gap by using *in-situ* electrochemical atomic force microscopy (EC-AFM) to directly observe the interfacial evolution during Zn electrodeposition and polymer adsorption on copper (Cu) substrates in the presence of varying concentrations of $ZnSO_4$ and PEO. Contrary to previous literature assumptions which emphasize the binding to the growing Zn crystal surfaces or $Zn^{2+}$ ions, our results demonstrate that PEO smooths Zn films by promoting nucleation of (002)-oriented Zn platelets through interactions with the Cu substrate. Density functional theory (DFT) simulations support this finding by showing that PEO adsorption on Cu modifies the interfacial energy of Zn/Cu/electrolyte interfaces, favoring the stabilization of Zn (002) on the Cu substrate, as well as confines Zn electrodeposition to a narrow near-surface region. These findings elucidate a novel design principle for electrode smoothing, emphasizing the importance of substrate selection paired with polymer additives that exhibit an attractive interaction with the substrate, but minimal interaction with growing crystals, offering a mechanistic perspective for improved battery performance.

## Introduction

The use of surfactants to modify interfacial free energies, enabling smooth interface growth, is well-established in vapor phase systems.[1–3] However, it remains uncertain how the concept works in solution-based electrochemical systems, which are essential across numerous technological fields – most notably in battery technologies, as well as in manufacturing processes of solution-based thin film deposition for organic solar cells, microelectronic devices, and even in geochemical systems where inorganic films heterogeneously form on mineral grains within reservoirs.[4–6] In the case of batteries, metal anodes such as zinc and lithium have attracted wide attention due to their high energy density but maintaining a flat interface has been a persistent challenge.[7–9] Most attempts have been Edisonian in nature, with a limited fundamental understanding of how to design these anodes to achieve a flat morphology that avoids dendrite formation resulting from these efforts. While several studies have explored the influence of electrolyte compositions on smoothing Zn or Li electrodeposits, and a few have investigated ways to control morphology by manipulating epitaxial relationships to promote the growth of flat-lying crystals, there remains a lack of comprehensive knowledge on how surfactants or polymer additives might impact or regulate electrode morphology.[10–19]

A previous finding suggests that the addition of a commonly used polymer additive, polyethylene oxide (PEO), can enhance the performance of Zn anode batteries, pointing to the potential of an interface smoothing mechanism similar to those observed in vapor-phase systems (**Fig. S1**).[20] From a synthetic perspective, one might expect PEO to adsorb onto various growing Zn crystal faces, thereby altering surface energies and facet-dependent growth rates to influence the Zn film morphology.[21,22] However, using real-time in situ observations of both interfacial evolution during Zn electrodeposition on Cu and the process of PEO adsorption on both the Cu and the Zn, we show that this hypothesis is incorrect. PEO does not interact with the growing Zn crystal surfaces to modify their morphology, rather it does indeed act as a surfactant by interacting specifically with the Cu substrate and modifying the interfacial solution structure. PEO-Cu binding modifies the interfacial free energy and confines ion distributions to a narrow interfacial zone, thereby promoting the nucleation of flat-lying (002)-oriented Zn platelets and preventing the random distribution of crystal orientations that typically cause surface roughening and eventual interface breakdown. Moving beyond previous concepts, such as polymer binding to $Zn^{2+}$ cations to adjust $Zn^{2+}$ distributions or to the growing Zn crystals to modify their surface energies, this study introduces a novel perspective and leads to a design principle for suppressing dendrite formation and mitigating interface roughening.[17,20,23–31] Specifically, it provides mechanistic insights into how substrate selection, coupled with systematic tuning of interfacial energy and structure via polymer additives that selectively bind to the substrate (without interacting with reactants or products), can be used to suppress dendrite formation in metal anode batteries.

## Results

**Zn electrodeposition on Cu substrates**

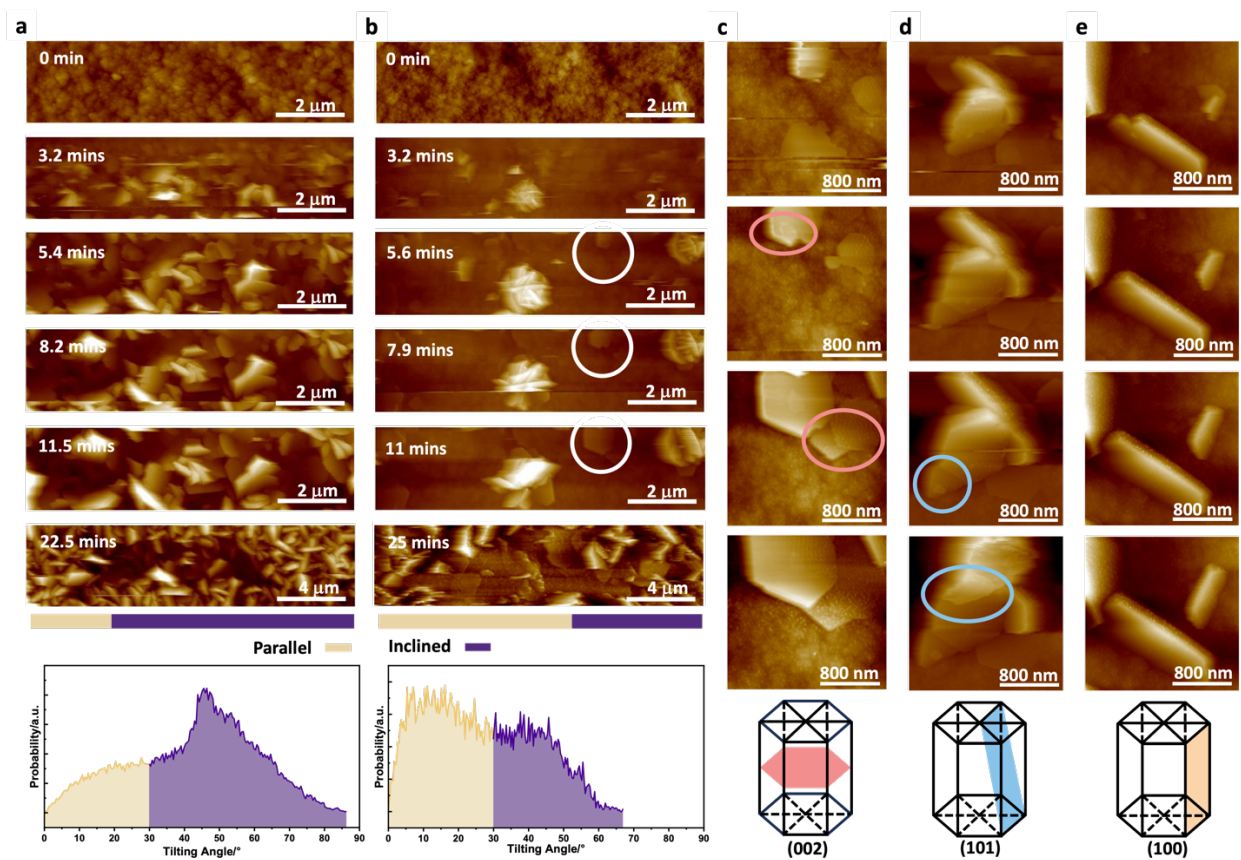

**Figure 1.** *In-situ* **EC-AFM results of Zn platelets electrodeposited on Cu in 0.1 M ZnSO$_4$ with and without PEO additives under constant voltage.** (a, b) The morphology evolution of Zn deposited on Cu in 0.1 M ZnSO$_4$ without PEO (a) and with 0.1 wt% PEO (b) (color range: 1.4 μm). Upper panels: time-series AFM images. Lower panels: distribution of the tilting angle of Zn platelets. Beige/purple sections in horizontal bars and in histograms in (a) and (b) represent parallel (tilting angle < 30°)/inclined plates (tilting angle >30°). White circles in (b) highlight flat-lying platelets. *In-situ* EC-AFM images show the growth of horizontal Zn plates with nucleation orientation of Zn (002) (c), inclined Zn plates with nucleation orientation of Zn (101) (d), vertical Zn plates nucleating as Zn (100) (e), maintaining a constant hexagonal sheet shape and aspect ratio. Red circles in (c) and blue circles in (d) highlight the layer-by-layer growth modes; the second and third layers will adopt the first layer as a template. Color range: 0.3 μm (c); 1.2 μm (d); 0.8 μm (e).

Using *in-situ* electrochemical atomic force microscopy (EC-AFM), we observed the electrodeposition of Zn from $ZnSO_4$ solutions on polycrystalline Cu substrates in pure and PEO-bearing solutions (**Fig. 1a, b**). In the absence of PEO, Zn platelets were observed to form with a range of crystallographic orientations leading to a rough interface (**Fig. 1a**). However, the results show that, in the presence of 0.1 wt% PEO (M.W. 100,000), the electrode surface remains much smoother due to a sharp increase in the ratio of parallel-to-inclined Zn plates. The morphology and statistical analysis of plate orientation for Zn electrodeposited in 0.1 M $ZnSO_4$ in the absence of PEO show that the Zn plates exhibit a wide range of orientations with 75% of the plates inclined at an angle >30° (**Fig. 1a**). In contrast, upon the addition of a small amount of PEO (0.1 wt%), the morphology was significantly altered. Although some inclined plates still formed, often in the form of a "desert rose", Zn plates oriented parallel to the Cu substrate were dominant. Analysis of the plate orientations shows that three are most common: flat-lying with Zn (002) parallel to the substrate, inclined such that Zn (101) forms the interface with the substrate, and vertical for which Zn (100) lies perpendicular to the substrate (**Fig. 1c, d, e**). Statistical analysis of the Zn plate orientations for the two cases shows that the ratio of Zn (002)-oriented to Zn (101)- and Zn (100)-oriented increases from 1:3 before adding PEO to 3:2 after (**Fig. 1a, b**).

For the Zn (002)-oriented plates, once the first layer of plates forms, subsequent layers (highlighted by red circles in **Fig. 1c**) use the first one as a template to continue growing plates that are parallel to the Cu substrate. The same is true of the inclined and vertical Zn plates; the second and third layers also adopt the first layer as a template and stack into multilayers (highlighted by blue circles in **Fig. 1d**). Thus, regardless of whether the first layer is composed of horizontal, inclined, or vertical Zn plates, the second and third layers will follow the orientation of the first layer and grow layer-by-layer, exhibiting the Frank-van der Merwe growth mode by which thin films grow layer-by-layer during vapor phase epitaxy at a low energy interface (**Fig. 1d, e, Fig. S2**).[32]

Although the orientation of the plates is altered in the presence of PEO, the sizes and aspect ratios of the Zn plates deposited at constant voltage are similar with and without PEO in solution (**Fig. 1a**, **b**), implying that PEO has no obvious effect on the relative growth kinetics of the Zn plates. Thus, if it interacts with the Zn crystal surfaces, the interactions are weak and neither adversely affects the delivery of $Zn^{2+}$ to the electrode surface, nor significantly alters Zn surface energies. In contrast, the increased ratio of the parallel-to-inclined Zn plates implies that PEO biases the orientation of the initial nuclei, which determines the orientations of the resulting plates. Thus, we hypothesize that PEO primarily interacts with the Cu substrate to adjust the interfacial structure of the Cu/electrolyte interfacial region and alter the interfacial energies of the Cu/Zn/electrolyte interface.

## PEO adsorption on Cu

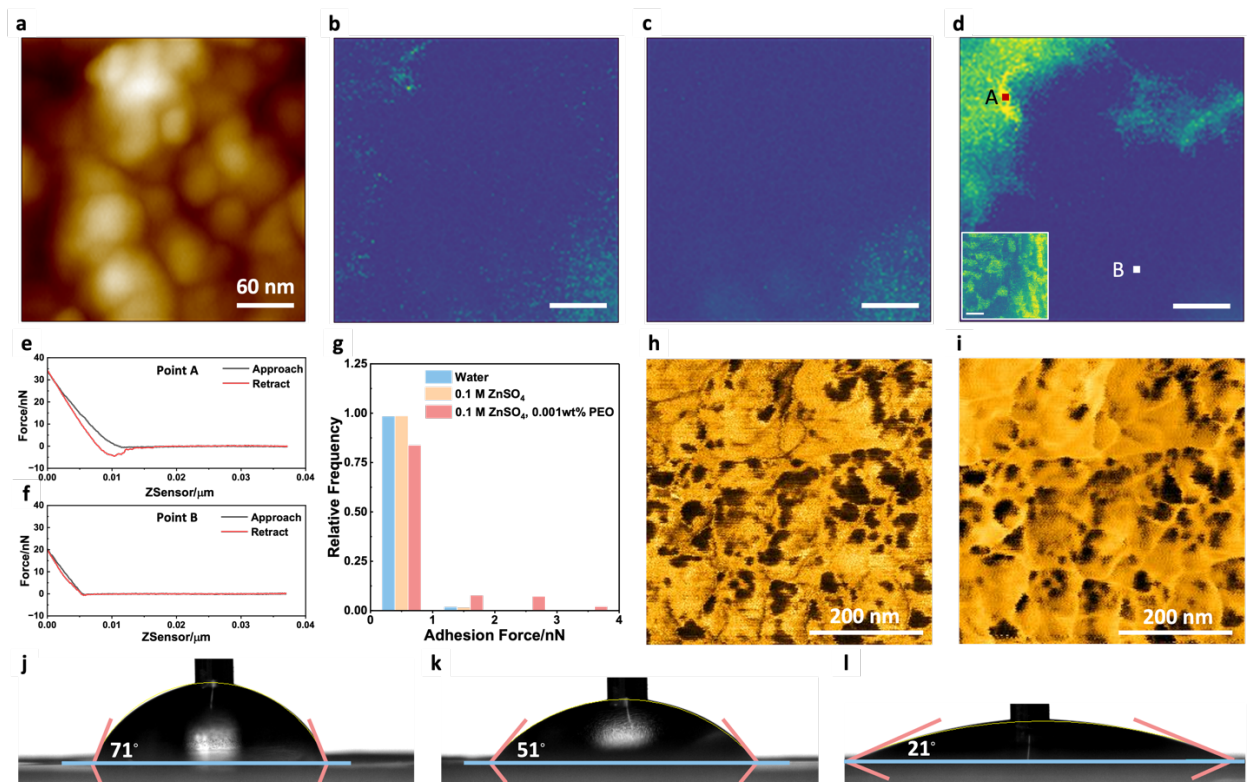

**Figure 2**. **Multiple approaches demonstrate PEO adsorption on Cu substrates**. (a) The AFM height image of the selected area on the Cu substrate under flowing conditions (color range: -15 to 15 nm). (b) The adhesion force map of the selected area shown in (a) in a water environment obtained by contact fast force mapping mode (color range from -0.67 to 3.83 nN). (c) The adhesion force map of the same selected region in (b) after flowing in 0.1 M $ZnSO_4$ solution until fully exchanged. (d) The adhesion force map of the same selected position in (c) after flowing in 0.001 wt% PEO, 0.1 M $ZnSO_4$ solution until 0.1 M $ZnSO_4$ solution is fully exchanged. Insert in (d): adhesion force map of Cu substrates in 0.1wt.% PEO solution (color range from -0.67 to 3.83 nN, scale bar 60 nm). (e, f) Single-point force curve of points A (e) and B (f) in panel (d) analyzed from adhesion force map in (d). (g) Relative frequency distribution of adhesion force on Cu substrates in water, and 0.1 M $ZnSO_4$ with and without 0.001 wt% PEO. (h) The PiFM signal image of the Cu substrate soaked in PEO solution collected at the PEO specific wavelength 1462 $cm^{-1}$. (i) Phase image collected at the same spot as the PiFM image in panel (h). (j-l) Water contact angle measurements of Cu substrates after treatment with PEO solutions of varying concentrations (0, 0.001 wt%, 0.1 wt%).

To test the hypothesis that PEO primarily interacts with the Cu substrate, Contact Fast Force Mapping (CFFM), photo-induced force microscopy (PiFM), and contact water angle measurements were conducted.[33–36] The adhesion force maps of the same region of the Cu substrate, which exhibited a granular morphology (**Fig. 2a**), were collected by CFFM in AFM under flowing conditions with and without PEO. The adhesion force map of the same region presented in **Fig. 2a** was collected in water and shows that the overall adhesion force is very weak and quite evenly distributed on the rough Cu surface (**Fig. 2b**). The adhesion force map was recollected in the same region after the solution was fully replaced by 0.1 M $ZnSO_4$ (**Fig. 2c**). The adhesion force remained weak and was also evenly distributed. The adhesion force map changed significantly once 0.1 M $ZnSO_4$ solution with 0.001 wt% PEO was introduced into the cell. The appearance of regions exhibiting higher adhesion force (lighter areas in **Fig. 2d**), demonstrates extensive polymer adsorption on the Cu surface.

To better interpret the CFFM data, single-point adhesion force curves were collected during CFFM at points within and away from the regions of adsorbed polymer seen in the CFFM images (see points A and B highlighted in **Fig. 2d**). The divergence of the approaching and retracting force curves at point A, which is within the region of polymer adsorption, is characteristic of a sticky surface. (As an AFM cantilever is retracted from a sticky surface, the probe remains adhered until the tensile force is large enough to cause a jump from contact, leading to a negative peak in the force curve). Based on the magnitude of the peak, we estimate an adhesion force of around 4 nN (**Fig. 2e**).[33,37] In contrast, the approaching and retracting force curves at point B, where no polymer is seen in the CFFM images, overlap well (**Fig. 2e**) demonstrating that point B is on the hard Cu surface where polymer adsorption is absent (**Fig. 2f**).[33,34]

From a statistical analysis of 16,384 force curves collected in three force maps over a 128 × 128 grid of points, we obtained the relative frequency distribution of adhesion forces on Cu substrates in the three liquid environments: water, 0.1 M $ZnSO_4$ solution, and 0.1 M $ZnSO_4$ solution with 0.001 wt% PEO (**Fig. 2g**). The results show that the addition of just 0.001 wt% PEO significantly alters the distribution of adhesion forces on Cu, resulting in approximately 20% of those forces exceeding 1 nN (**Fig. 2g**). When the PEO concentration was increased to 0.1 wt% in the solution, which is equal to the concentration used in the Zn electrodeposition experiments, the proportion of adhesion forces exceeding 1 nN reached 80% (**Fig. 2d inserted and Fig. S3**). Both height and phase images of Cu substrates soaked in PEO solution were collected to visualize the process of PEO adsorption on Cu substrates, which further supported the conclusion that PEO adsorbs on Cu (**Fig. S4, 5, and SI text**).

Chemical mapping and surface properties of Cu substrates after treatment with PEO solutions provide further evidence that PEO adsorbs to the Cu. Using dried Cu-PEO samples (see methods in the **SI text** for details on sample preparation), we performed PiFM, which is a technique that combines AFM and infrared spectroscopy, to obtain nanoscale chemical maps of PEO on Cu substrates.[35,36] The PiFM and phase images of Cu-PEO samples were collected simultaneously at an excitation wavelength 1462 cm$^{-1}$, which is the specific excitation wavelength of PEO.[38,39] The bright regions in the PiFM image indicate the chemical signal of PEO (**Fig. 2h**). Moreover, the bright regions in the phase image correspond to a softer material (**Fig. 2i, SI text**). Because the pattern of bright regions in the PiFM and phase images are similar, we can conclude that soft, sticky regions seen in phase or CFFM measurements consist of PEO on Cu.

To evaluate the extent to which the interfacial energy of the Cu surface is affected by PEO adsorption, contact angles of water droplets were measured on bare Cu and Cu subject to treatment with different concentrations (0.001wt%, 0.1wt%) of PEO solution. The results show that the wettability of Cu substrates can be greatly improved after soaking in PEO solution. The contact angle for bare Cu after three minutes of plasma treatment was 71° (**Fig. 2j**). On plasma-treated Cu substrates soaked in PEO solution for 2 hours and then dried (for the sample preparation protocol, see the **SI Methods**), the water contact angle was 51° and 21° for 0.001 wt% and 0.1 wt% PEO solutions, respectively (**Fig. 2k, l**).

**Zn oxidation and PEO adsorption on Zn**

Next, we show that the interaction between PEO and the surface of Zn is too weak to control the overall morphology of the Zn films. To investigate the adsorption behavior of PEO on Zn metal, Zn metal was first electrodeposited on Cu before being subjected to PEO adsorption. However, once the voltage that drives electrodeposition was turned off, this composite substrate became unstable in water as Zn plates exhibited etching and surface reorganization (**Fig. S6**). This effect is attributed to Zn oxidation.[40–42] Both Transmission Electron Microscopy (TEM) and Grazing Incidence X-ray Diffraction (GIXRD) results showed that the resulting layered hexagonal structures on Cu were oxidation products of Zn metal (**Fig. S7 and S8**).

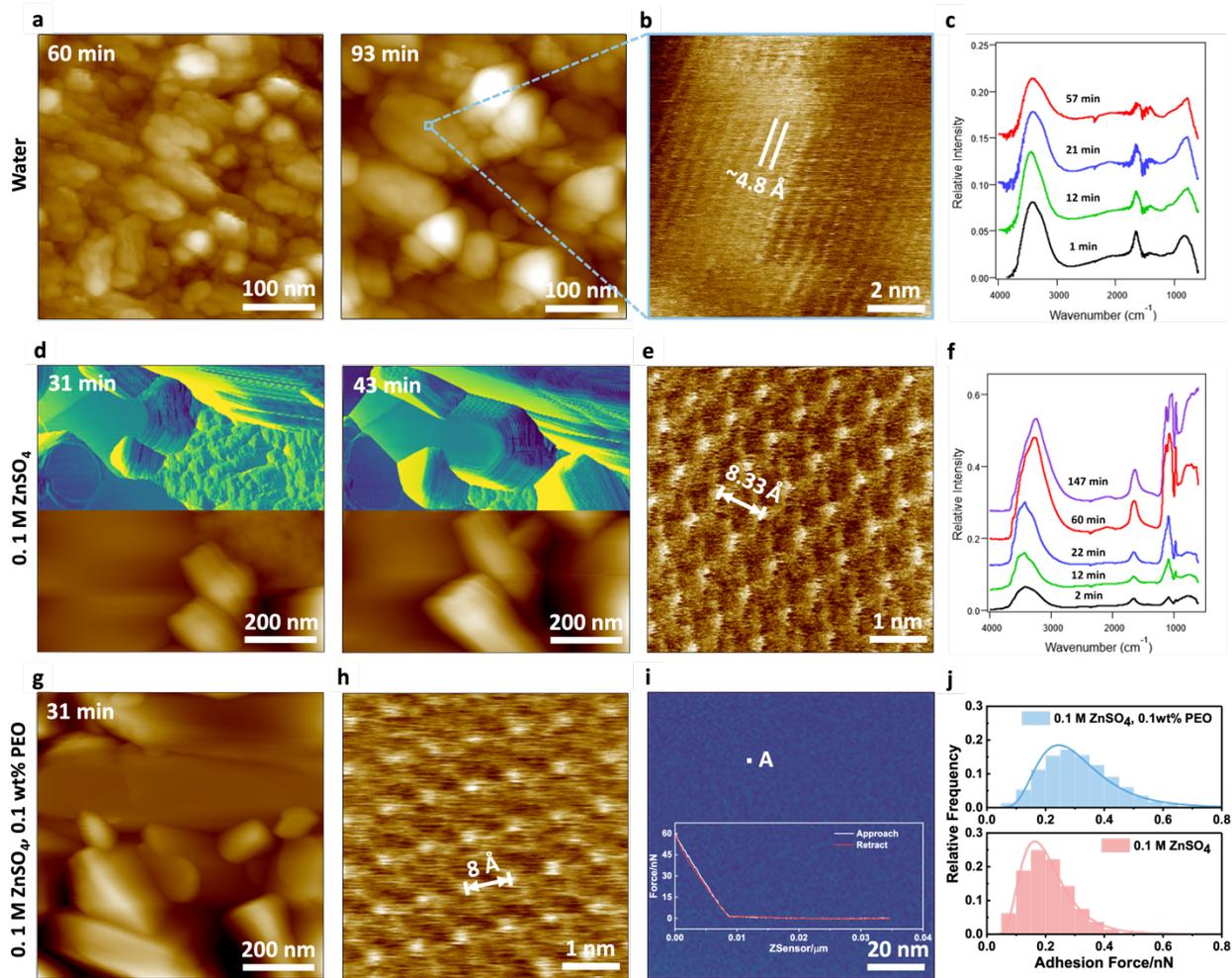

**Figure 3. Surface reorganization and oxidation of Zn foil in water and ZnSO₄ with and without PEO additives.** (a) AFM height images of the Zn metal after soaking in water for 60 and 93 minutes (color range: 125 nm). (b) The zoomed-in height image of (a) presents the layer spacing (color range: 450 pm). (c) *In-situ* ATR-FTIR of Zn in water. For clarity, the baseline of the spectra is offset by a constant at increasing reaction times. (d) The phase (top part) and height (bottom part) images of Zn metal after soaking in 0.1 M ZnSO₄ solution for 31 and 43 minutes (color range of the height image: 280 nm). (e) The zoomed-in height image of the hexagonal layered structure in (d) with lattice-resolved resolution (color range: 300 pm). (f) *In-situ* ATR-FTIR of Zn in 0.1 M ZnSO₄ solution. For clarity, the baseline of the spectra is offset by a constant at increasing reaction times. (g) The height image of Zn metal after soaking in 0.1 M ZnSO₄ solution with 0.1 wt% PEO for 31 minutes (color range: 280 nm). (h) The zoomed-in height image of the hexagonal layered structure in (g) with lattice-resolved resolution (color range: 200 pm). (i) Adhesion force map of Zn metal in ZnSO₄ solution with 0.1 wt% PEO. Insert in (i): The single-point force curve of point A analyzed from the adhesion force map in (i). (j) Relative frequency distribution of adhesion force on Zn substrates in 0.1 M ZnSO₄ with and without 0.1 wt% PEO.

To delve into the nature of surface reorganization of the Zn, *in-situ* AFM and *in-situ* attenuated total reflectance Fourier transform infrared spectroscopy (ATR-FTIR) were conducted on commercially available Zn foils in water (pH=7) and in 0.1 M ZnSO₄ (pH=5.5) with and without

0.1 wt% PEO. Commercially available Zn foil initially exhibits a granular morphology in water (**Fig. 3a and Fig. S9a**). However, after 33 minutes, the grains grow larger until they finally form bundles of platelets (**Fig. 3a, Fig. S9b, Figure S10, Video S1**). After two hours of immersion in water, Zn foils show visible pitting and a greyish film (**Fig. S11**). Higher resolution images of the newly formed structure in **Fig. 3a** captured in situ exhibit a lattice spacing expected for $Zn(OH)_2$ (**Fig. 3b and Fig. S12**).[43]

The *in-situ* ATR-FTIR measurements were conducted to test Zn chemical transformation in water (**Fig. 3c**). Upon introduction of water on the zinc foil surface, the ATR-FTIR spectra showed a prominent OH stretch vibration at 3410 $cm^{-1}$ with a shoulder band at ca. 3220 $cm^{-1}$, along with the water bending vibration at 1658 $cm^{-1}$ and a strong peak at 834 $cm^{-1}$. The latter has been previously reported by Srivastava and Secco[44], Geetha and Sakthi[45] and Wang and Andrews[46] and attributed to a ZnOH bending and OH twisting vibration. These results clearly point to the formation of a hydrated $Zn(OH)_2$ surface. As the hydrolysis reaction progressed, both the stretch and bending vibration peaks dropped in intensities while that of the 834 $cm^{-1}$ remained constant, suggesting the growth of the $Zn(OH)_2$ layer and the decrease of the amount of hydration water within the detection depth. The much more pronounced reduction of the water bending vibration peak further confirmed the contribution of the hydroxyl groups in the $Zn(OH)_2$ phase.

When Zn foil was immersed in 0.1 M $ZnSO_4$ solution, the *in-situ* AFM results show that the Zn reorganized rapidly into a layered hexagonal structure, which replaced the initial granular structures until they fully occupied the surface (**Fig. 3d** and **Fig. S13**, **Fig. S14, Video S2**). High-resolution images show that the lattice is the same as that of $Zn_4SO_4(OH)_6 \cdot xH_2O$ (ZHS, x=4), with lattice parameters a and c equal to 0.833 nm (**Fig. 3e, Fig. S15**).[47,48]

*In-situ* ATR-FTIR measurements were conducted to determine whether the proposed Zn chemical transformation in 0.1 M $ZnSO_4$ solution was indeed occurring (**Fig. 3f**). Upon the introduction of 0.1 M $ZnSO_4$ solution, dramatic changes in the ATR-FTIR spectra were observed. In the OH stretch vibration region, the ATR-FTIR peaks increased in intensity and quickly became more structured as the reaction progressed and the spectral maximum red-shifted from 3421 $cm^{-1}$ two minutes after solution introduction to 3241 $cm^{-1}$ after reacting for 147 minutes (**Fig. 3f**). Similarly, the water bending peak increased in intensity while also red-shifted from 1654 $cm^{-1}$ to 1628 $cm^{-1}$. More pronounced spectral changes appeared in the spectral region of < 1300 $cm^{-1}$, mostly corresponding to the $SO_4^{2-}$ vibrational modes. A single peak at 1079 $cm^{-1}$ at the beginning of the reaction (2 min) greatly increased in intensity while further split to four peaks at 1026 $cm^{-1}$, 1056 $cm^{-1}$, 1120 $cm^{-1}$, and 1162 $cm^{-1}$, respectively, at longer reaction times. The peak at 967 $cm^{-1}$ gained intensity greatly while gradually red-shifted from 961 cm$^{-1}$. There were also large intensity gains for peaks at < 900 cm$^{-1}$.

The free SO$_4^{2-}$ anion possesses a T$_d$ symmetry and normally only the anti-symmetric stretch vibration (ν$_3$) at ~ 1120 cm$^{-1}$ and anti-symmetric bending vibration (ν$_4$) at ~ 650 cm$^{-1}$ are observed on the infrared spectra.[49] However, as the SO$_4^{2-}$ anion coordinates to metal ions, the T$_d$ symmetry is lowered depending on the specific binding patterns and as a consequence, not only the symmetric stretch (ν$_1$) at ~ 1020 cm$^{-1}$ and symmetric bending (ν$_2$) at ~ 460 cm$^{-1}$ become partially allowed on the infrared spectra, spectral splitting also occur as dictated by the corresponding symmetries.[49–52] Both the appearance of three or more peaks within the range of the ν$_3$ vibrational mode and the relative peak positions observed in this work (**Fig. 3f**) are consistent with the formation of Zn$_4$SO$_4$(OH)$_6$•xH$_2$O (ZHS) mineral phases with varied hydration waters.[49–51] The continuously evolving spectral patterns in the regions of the OH stretch, water bending and the SO$_4^{2-}$ vibrational modes suggested that hydration states and the relative populations of the mineral phases varied on the time scale of the experiments. Particularly, the red-shifted, well-structured OH stretch vibration peaks in the range of 3200 cm$^{-1}$ to 3500 cm$^{-1}$ indicated the transformation of ZHS phases from weak and less-ordered H-bonding environment to one that is strong and well-ordered. The increasing water bending band at 1628 cm$^{-1}$ clearly points to the presence of hydration waters in the formed minerals at all the reaction stages.

When Zn foil was immersed in 0.1 M ZnSO$_4$ solution with 0.1 wt% PEO, after half an hour the surfaces were again fully reorganized into the hexagonal layered structure (**Fig. 3g, Fig. S16**) with a lattice that perfectly matched that of ZHS (**Fig. 3h, Fig. S17**), demonstrating that PEO does not hinder the surface reorganization or chemical transformation of Zn. Also, the adhesion force map collected on the hexagonal layered structure formed in 0.1 M ZnSO$_4$ with or without PEO is almost identical: no bright regions were observed, demonstrating that no PEO polymer adsorbed onto the Zn surfaces after oxidation (**Fig. 3i, Fig. S18**). The single-point force curve at point A showed the typical force curve for the hard Cu surfaces (**Fig. 3i**) and the distribution of adhesion forces across 16,384 points on Zn foils in 0.1 M ZnSO$_4$ with and without 0.1 wt% PEO was uniformly below 1 nN (**Fig. 3j**), confirming a lack of polymer adsorption. However, the average adhesion force in 0.1 M ZnSO$_4$ increased by roughly 0.1 nN upon addition of 0.1 wt% PEO, indicating a small increase in the nearly negligible adhesion force due to PEO in the solution or adsorbed to the tip (**Fig. 3j**).

The above results imply that multiple chemical reactions and physical processes are happening in the Zn-aqueous battery system, including oxidation of Zn or reduction of Zn$^{2+}$ reactions and polymer adsorption or desorption on Zn or ZHS. The correlated *in-situ* height and phase images demonstrate the adsorption of PEO on Zn and the desorption of PEO during surface reorganization

(**Fig. S19**). PEO adsorption on the Zn surface cannot outcompete the natural and spontaneous oxidation of the Zn surface, indicating a weak interaction between Zn and PEO. However, the Zn electrodeposition process is driven by a significant external electrochemical driving force, enabling this reduction reaction to effectively overcome the naturally occurring oxidation of Zn. Consequently, PEO adsorption does not influence Zn electrodeposition, which constitutes the growth process in our system.

To summarize, the above results show that PEO smooths the electrodeposited Zn on Cu electrodes because PEO adsorbs on the Cu electrodes and strongly biases the orientation of the Zn nuclei electrodeposited on the Cu surface. In contrast, PEO weakly interacts with the Zn surface and does not influence the Zn growth process. Once Zn is electrodeposited on the Cu surface, the weak Zn-PEO binding does not prevent Zn from oxidizing in the absence of a reducing electric field (E-field), causing any adsorbed PEO to detach from the oxidized Zn.

**Mechanism by which PEO biases the nucleation orientations**

The above experimental findings lead us to hypothesize that the adsorption of PEO on the Cu surface influences the local interfacial structure and energy around regions of PEO adsorption, thus biasing the orientation of Zn nuclei. To test this hypothesis, we used simulations to quantify the effect of PEO on three aspects of Zn nucleation on the Cu: 1) biasing the facet specificity of the interfacial energy for Zn nucleation on Cu, 2) altering the structure of the interfacial electrolyte layer, and 3) the dynamics of $Zn^{2+}$ ions at the interface. Density functional theory (DFT) was employed to compare the role of PEO in biasing the thermodynamics of orientational specificity of Zn nucleation.[53–55] Vacuum calculations revealed the energetic differences for the nucleation of Zn (100) and Zn (002) faces on the Cu (111) surface. The Zn nanoparticle bound with two (100) and six (002) faces was brought into contact with the Cu surface, forming either direct Cu (111) / Zn (100) or Cu (111) / Zn (002) contact. It was found that the interfacial energy of Cu / Zn (100) is 1.2 times lower than that of Cu / Zn (002), demonstrating the thermodynamic preference for nucleation of Zn (100) instead compared to Zn (002) faces (**Fig. S20**).

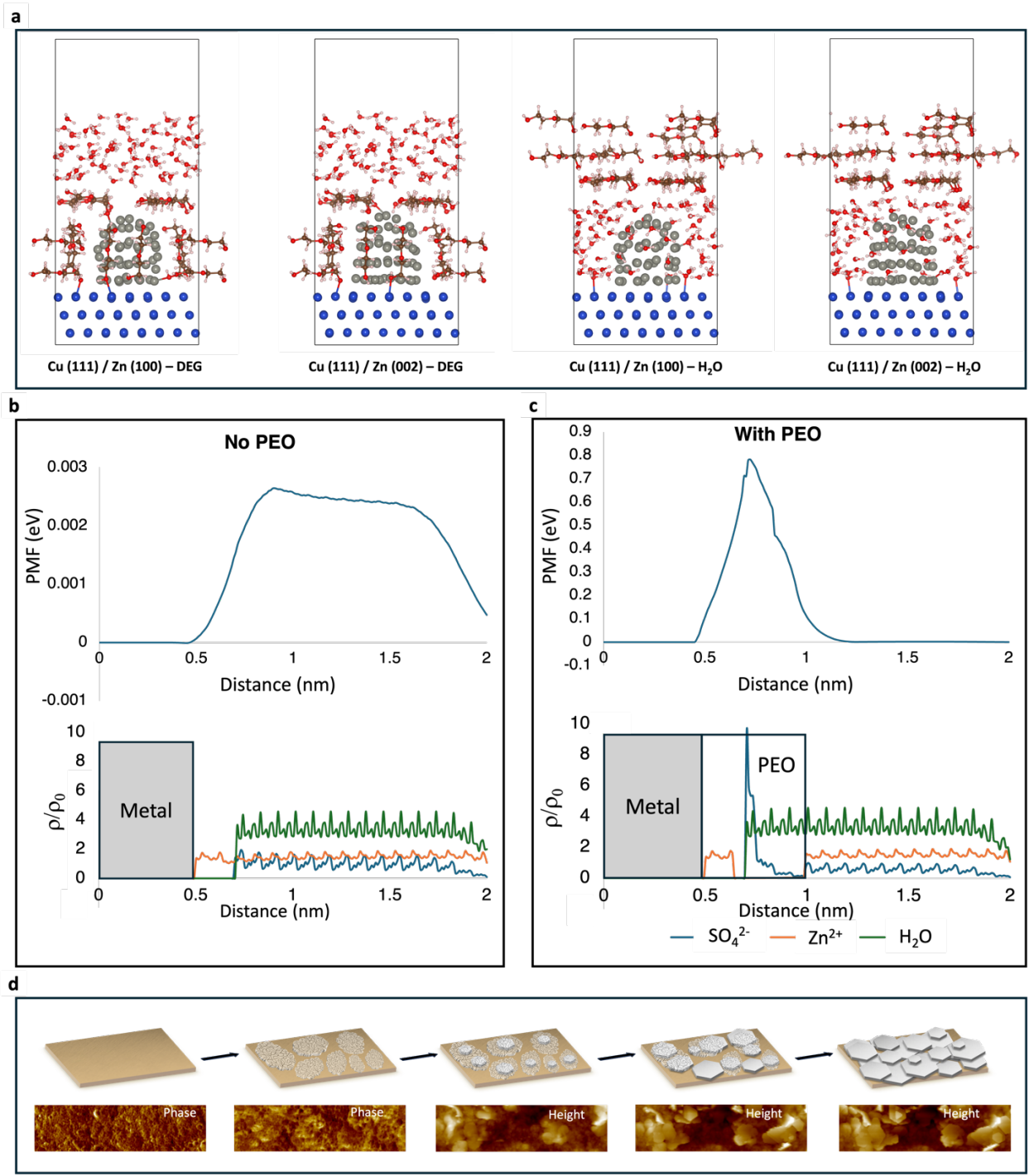

**Figure 4. Mechanism of how PEO biases the nucleation orientations**. (a) The relaxed structures from DFT energy calculations of Cu (111) / Zn (100) and Cu (111) / Zn (002) interfaces in diethylene glycol (DEG)- and $H_2O$-saturated environments. These structures are shown in ball-and-stick form, with Cu, Zn, O, C, and H atoms represented in blue, gray, red, brown, and white, respectively. (b, c) The cDFT simulation results of Cu/electrolyte interfacial structures in $ZnSO_4$ solution with and without PEO. Relative water density was scaled by a factor of 3 for clarity. (d)

Schematic illustration correlated with AFM images showing that PEO creates a barrier to the electrodeposition of vertically-growing Zn platelets and promotes flat-lying layer-by-layer growth.

**Table 1. Comparison of energies of Cu / Zn (100) and Cu / Zn (002) interfaces in H$_2$O and PEO (DEG) saturated environments. The energy difference $\Delta E$ is defined as $\Delta E = E[\text{Cu}(111)/\text{Zn}(100)] - E[\text{Cu}(111)/\text{Zn}(002)]$.**

| Solvent | $\Delta E$ (kJ/mol) |
|---|---|
| PEO (DEG) | 294.2 |
| H$_2$O | -369.9 |

The effect of solvent and PEO on interfacial energy was calculated using the comparison of total energies of solvated interfaces. The DFT simulations of Cu (111) / Zn (100) and Cu (111) / Zn (002) were repeated in water- and PEO- environments (**Fig. 4a**). To keep the same number of each species constant, simulations were performed with 74 water molecules and 14 Diethylene glycol (DEG) molecules representing PEO monomers. Following the vacuum simulations, we considered two orientations of Zn nanoparticles on Cu surfaces and two solvation modes, i.e., water molecules or DEG molecules in direct contact with Zn and Cu surfaces. The packing arrangements of Zn (100) and Zn (002) on Cu (111) remained similar after relaxation in the two modes of solvation (**Fig. S21** and **Table S1**).

Our DFT results demonstrate that PEO (DEG) adsorption on Cu modifies the interfacial energy of Zn/Cu/electrolyte interfaces, favoring the stabilization of Zn (002) on the Cu substrate. Specifically, while Zn (100) on Cu is thermodynamically favored in water-saturated conditions, the presence of PEO stabilizes Zn (002) on Cu as the energetically preferred orientation (**Table 1**). Consequently, PEO-saturated regions promote the formation of flat-lying Zn (002) plates parallel to the Cu substrate, while water-saturated regions result in vertical Zn (100) plates oriented perpendicularly. These results align well with our experimental observations, which show a mixture of plates in various orientations but with a clear bias in orientation distribution after the addition of PEO.

Thermodynamic preference of certain orientations of the growing particles is not necessarily insufficient to understand the interfacial processes occurring during electrodeposition, in which kinetics often drives the formation of high-energy morphologies.[56] Coupled kinetic Poisson-Nernst-Planck (PNP) and classical DFT (cDFT) simulations were conducted to elucidate how PEO affects electrodeposition kinetics and identify the balance of thermodynamic and kinetic factors controlling the orientation of Zn plates. These simulations provide insights into the driving forces for the deposition of Zn$^{2+}$ ions onto the electrode surface and the role of the structure and dynamics in the electric double layer at the electrode surface on Zn$^{2+}$ mobility and dynamic distribution. Simulations of Zn$^{2+}$ potential of mean force (PMF) at an electrified electrode at aqueous electrolyte interfaces predict barrier-free electrodeposition (**Fig. 4b**). There is a weak repulsion between $Zn^{2+}$ ions and the electrode surface at the distances of 0.75 – 1.75 nm from the electrode surface. However, the repulsive PMF is smaller than 0.0025 eV or about ten times smaller than the thermal energy at room temperature and, thereby, does not affect the electric-potential-driven deposition. Under these conditions, the $Zn^{2+}$ supply to the surface is unimpeded, and electrodeposition is a reaction-limited process. The reaction-limited electrodeposition drives the formation of thermodynamically stable morphologies,[56] and the formation of plates in (002) orientation to the surface is predicted.

In contrast, in the presence of PEO, which creates a low dielectric layer at the interfaces, the PMF of $Zn^{2+}$ ions has a barrier of 0.8 eV (**Fig. 4c**). In these simulations, the polymer layer was modeled as a uniform film with a dielectric constant of 2. It was permeable for the ions but created a distinct dielectric environment at the boundary between the metal electrode and aqueous solution with the dielectric constant of 78.5 to simulate PEO adsorption at the interface (**Fig. 4c**). The corresponding steady-state distribution of $Zn^{2+}$ ions has a depletion zone 0.7 -1 nm from the surface due to counter ions accumulation about 0.75 nm from the interface (**Fig. 4c**). The significant barrier in the PMF and partial depletion of $Zn^{2+}$ ions close to the interface suggest that PEO shifts the mechanism of electrodeposition toward diffusion-limited process. Under these conditions, either higher energy morphologies develop when reaction- and diffusion-driven mechanisms compete, or dendrites are formed in a purely diffusion-limited process.

Specifically, if at the pure electrode/aqueous electrolyte interface, the reaction kinetics is controlled by the reaction rate, in the presence of PEO electron transfer rate and diffusion rate of $Zn^{2+}$ ions both define the growth rate and morphology of the electrodeposited Zn.[57,58] When the reaction rate of $Zn^{2+}$ ions reducing to Zn is relatively larger than the $Zn^{2+}$ diffusion rate, the whole process will be diffusion-limited due to a slow diffusion rate, and only $Zn^{2+}$ in near-surface regions can be reduced to Zn due to a lack of constant supply of $Zn^{2+}$ ions in regions slightly away from the surface. In this way, PEO adsorption confines electrodeposition to a narrower interfacial region, reduces the probability of vertical and inclined growth, and promotes parallel orientation of Zn plates. The parallel orientation is also thermodynamically more favorable in the presence of PEO, as the energy simulations predicted (**Table 1**). Thereby, a higher fraction of parallel-orientated Zn plates is expected under mixed reaction-diffusion conditions. Reducing the diffusion rate by lowering $ZnSO_4$ concentration shifts electrodeposition toward a diffusion-limited process, leading to Zn plates with a bigger parallel-to-inclined ratio (**Fig. S22**). Similarly, modulating the relative rate of reaction and diffusion processes and increasing the reaction rate by increasing the current density can also explain the results of other electrodeposition studies. Specifically, applying a high current density below a certain threshold has been shown to produce a planar structure composed of non-epitaxial Zn (002)-oriented platelets.[59–61]

By integrating *in-situ* measurements with computational simulations, the above findings provide a clear mechanistic understanding of how a specific polymer, PEO flattens the electrode surface: PEO first adsorbs onto the Cu substrate, then Zn metal nucleates as Zn (002) beneath the PEO, followed by PEO detachment as the Zn nuclei grow. Subsequently, the Zn platelets grow layer-by-layer using the first layer of Zn (002) as a template (**Fig. 4d**)

## Discussion

Multiple side reactions accompany Zn electrodeposition, including hydrogen evolution and oxygen reduction, which increase the pH. This pH shift leads to the precipitation of $Zn(OH)_2$, ZHS, and ZnO from the supersaturation solution, both of which also present plate-like morphologies.[11] Thus, the plates observed in EC-AFM might be Zn reduced from $Zn^{2+}$ under a reducing E-field, or $Zn(OH)_2$, ZHS and ZnO precipitated due to the pH increase. To test for this phenomenon, we hindered the oxygen reduction reactions by removing dissolved oxygen from the solution. However, a similar evolution in morphology was observed with and without dissolved oxygen, indicating that the platelet formation is not attributable to side reactions related to oxygen reduction (**Fig.1 and Fig S23**). To investigate whether the increase in pH can generate the observed platelets, liquid AFM under flowing conditions and increasing pH was conducted. When the pH exceeds 6.3, 6.6 and 6.8, 0.1 M $ZnSO_4$ solutions become supersaturated with respect to ZHS, ZnO (Zincite), and $Zn(OH)_2$, respectively. However, we do not observe any morphological changes or formation of platelets on the Cu surface in conditions of pH 6.6, 7.2, or 8 (**Fig. S24**), demonstrating that the expected pH increase is not sufficient to drive the formation of $Zn(OH)_2$, ZnO or ZHS on Cu. Thus, the hexagonal platelets we observe in *in-situ* EC-AFM must be Zn reduced from $Zn^{2+}$. The $Zn(OH)_2$ and ZHS observed in ex-situ experiments (**Fig. S7**, **S8, and Fig. S25**) both in this study and others likely comes from the oxidation of Zn into $Zn(OH)_2$ and ZHS after the reducing E-field is turned off,[15,40,42,62] though we cannot rule out the possibility that, in some studies, $Zn(OH)_2$ and ZHS crystals formed in bulk solution and deposited on the surface.

In addition to side reactions, the influence of potential alloying of Zn into Cu was considered but was excluded, as no difference was observed for the electrodeposition of Zn on fresh Cu or on Cu that was cycled until reaching the steady state under the same experimental conditions. Furthermore, the influence of Cl⁻ impurities in $ZnSO_4$ salts was considered but excluded due to the trace amounts at which they were present (1-2 μM in 0.1 M $ZnSO_4$ solution).

Another hypothesis was considered to explain why PEO influences the orientations of the Zn nucleus on Cu substrates. One commonly observed scenario involves preferential adsorption of ligands on particular Cu facets, leaving those Cu facets that are less favorable to PEO adsorption exposed. Thus grains of the substrates that present specific crystallographic orientations might facilitate the nucleation of Zn (002) plane.[15] Electron backscatter diffraction (EBSD) and Scanning Electron Microscopy (SEM) measurements were conducted to test this hypothesis.[63] The orientation map of the Cu substrate before Zn electrodeposition combined with SEM images of the Zn morphology deposited on the same Cu spot after 30 seconds, shows no clear correlation between the crystallographic orientations of Zn plates and the grains of the underlying Cu substrate (**Fig. S26**). These data show face-selectivity of PEO adsorption to Cu substrates is not responsible for controlling the orientation of Zn plates during nucleation.

## Conclusions

This study demonstrates a distinct mechanism for using polymers to smooth electrodes in Zn aqueous battery systems and answers the long-lasting question of why a widely used polymer, PEO, smooths electrodes, introducing a novel design principle for interface smoothing in solution-based electrochemical systems. This approach leverages polymer additives with appropriately strong binding to the substrate and weaker binding to the deposited metal to precisely tune interfacial energy and structure, enabling control over the orientations of nuclei in the initial thin layer of Zn and promoting continued layer-by-layer growth.

Moreover, by using such developed platforms and standardized protocols, extending the testing timeframe beyond initial electrodeposition to multiple cycles will deepen our understanding of how polymer additives affect long-term battery cycling performance. This principle can be further refined by exploring how electrolyte composition and polymer chemical structure impact polymer-induced electrode smoothing in electrochemical systems.

Rational design of the electrode surface can be accomplished by grafting single polymer layers on the electrode surfaces, which is an efficient alternative for conventional surface modification, such as surface coating. This approach achieves interface smoothing by forming polymer-saturated regions in solution environments, all while maintaining electron conductivity and reducing the high costs typically associated with conventional surface modification. It offers an efficient and cost-effective strategy to enhance device stability, performance, and longevity across a wide range of applications where precise control of the electrode interface smoothness is critical, including batteries, electrocatalysis, solar cells, corrosion protection, and microelectronics.

## Acknowledgments

The authors acknowledge Prof. John C. Berg's group for their assistance with water contact angle measurements, and Dr. Thi Kim Hoang Trinh for the $N_2$ bubbling setup used for oxygen removal in solution. They also express gratitude to Dr. Mitchell Kaiser, Dr. Benjamin A. Legg, and Dr. Elias Nakouzi for their valuable discussions. This work was supported by the U.S. Department of Energy (DOE) Office of Science, Basic Energy Sciences, Materials Sciences and Engineering Division at PNNL (FWP 12152). The AFM and SEM were conducted at the Pacific Northwest National Laboratory and the University of Washington Molecular Analysis Facility (MAF). MAF is a National Nanotechnology Coordinated Infrastructure site at the University of Washington supported in part by the National Science Foundation (grant NNCI-1542101), the University of Washington, the Molecular Engineering & Sciences Institute, and the Clean Energy Institute. The PiFM was conducted at the University of Washington Molecular Analysis Facility. TEM imaging, FTIR, and XPS were performed in the Environmental and Molecular Sciences Laboratory, a Department of Energy Office of Science User Facility at PNNL sponsored by the Office of Biological and Environmental Research. Simulations were performed using the resources of the National Energy Research Scientific Computing Center, a DOE Office of Science User Facility supported by the Office of Science of the U.S. Department of Energy under Contract No. DE-AC02-05CH11231 using NERSC award BES-ERCAP0028725. Work by CO was performed under the auspices of the U.S. Department of Energy by Lawrence Livermore National Laboratory under Contract DE-AC52-07NA27344. J.S.D. acknowledges a Washington Research Foundation Postdoctoral Fellowship. PNNL is a multi-program national laboratory operated for the Department of Energy by Battelle under Contracts No. DE-AC05-76RL01830.

**Author Contributions:** Y.X., J.J.D.Y, and J.L. designed research; Y.X., D.S., M.Z., Z.W., C.S., J.S.D., S.H.R.S., M.H.E., J.T., and M.L.S. performed research; Y.X., M.Z., Z.W., P.K.T., C.A.O., and J.J.D.Y analyzed data; and Y.X., D.S., Z.W., M.L.S., J.J.D.Y., and J.L. wrote the paper.

**Competing interests:** The authors declare no competing financial interests.

**Data and materials availability:** All data are available in the main text or the supplementary materials.

# References
[1] D. Kandel, E. Kaxiras, in *Solid State Physics* (Eds.: H. Ehrenreich, F. Spaepen), Academic Press, **2000**, pp. 219–262.
[2] M. Copel, M. C. Reuter, E. Kaxiras, R. M. Tromp, *Phys. Rev. Lett.* **1989**, *63*, 632–635.
[3] M. Copel, M. C. Reuter, M. Horn von Hoegen, R. M. Tromp, *Phys. Rev. B* **1990**, *42*, 11682–11689.
[4] S. Bai, Z. Huang, G. Liang, R. Yang, D. Liu, W. Wen, X. Jin, C. Zhi, X. Wang, *Advanced Science* **2024**, *11*, 2304549.
[5] L. P. Colletti, B. H. Flowers, J. L. Stickney, *Journal of The Electrochemical Society* **1998**, *145*, 1442.
[6] S. S. Zhang, *Journal of Power Sources* **2006**, *162*, 1379–1394.

[7] H. Pan, Y. Shao, P. Yan, Y. Cheng, K. S. Han, Z. Nie, C. Wang, J. Yang, X. Li, P. Bhattacharya, K. T. Mueller, J. Liu, *Nature Energy* **2016**, *1*, 16039.
[8] X. Yuan, B. Liu, M. Mecklenburg, Y. Li, *Nature* **2023**, *620*, 86–91.
[9] J. Xiao, *Science* **2019**, *366*, 426–427.
[10] T. P. Moffat, D. Wheeler, D. Josell, *Journal of The Electrochemical Society* **2004**, *151*, C262.
[11] S. Peulon, D. Lincot, *Journal of The Electrochemical Society* **1998**, *145*, 864.
[12] J. S. Keist, C. A. Orme, P. K. Wright, J. W. Evans, *Electrochimica Acta* **2015**, *152*, 161–171.
[13] X. Yang, Z. Dong, G. Weng, Y. Su, J. Huang, H. Chai, Y. Zhang, K. Wu, J.-B. Baek, J. Sun, D. Chao, H. Liu, S. Dou, C. Wu, *Advanced Energy Materials* **2024**, *14*, 2401293.
[14] J. Zheng, L. A. Archer, *Science Advances* **n.d.**, *7*, eabe0219.
[15] L. Ren, Z. Hu, C. Peng, L. Zhang, N. Wang, F. Wang, Y. Xia, S. Zhang, E. Hu, J. Luo, *Proceedings of the National Academy of Sciences* **2024**, *121*, e2309981121.
[16] J. Zheng, Q. Zhao, T. Tang, J. Yin, C. D. Quilty, G. D. Renderos, X. Liu, Y. Deng, L. Wang, D. C. Bock, C. Jaye, D. Zhang, E. S. Takeuchi, K. J. Takeuchi, A. C. Marschilok, L. A. Archer, *Science* **2019**, *366*, 645–648.
[17] S. Jin, J. Yin, X. Gao, A. Sharma, P. Chen, S. Hong, Q. Zhao, J. Zheng, Y. Deng, Y. L. Joo, L. A. Archer, *Nature Communications* **2022**, *13*, 2283.
[18] P. Biswal, S. Stalin, A. Kludze, S. Choudhury, L. A. Archer, *Nano Lett.* **2019**, *19*, 8191–8200.
[19] G. Bergman, N. Bruchiel-Spanier, O. Bluman, N. Levi, S. Harpaz, F. Malchick, L. Wu, M. Sonoo, M. S. Chae, G. Wang, D. Mandler, D. Aurbach, Y. Zhang, N. Shpigel, D. Sharon, *J. Mater. Chem. A* **2024**, *12*, 14456–14466.
[20] Y. Jin, K. S. Han, Y. Shao, M. L. Sushko, J. Xiao, H. Pan, J. Liu, *Advanced Functional Materials* **2020**, *30*, 2003932.
[21] Z. Chen, T. Balankura, K. A. Fichthorn, R. M. Rioux, *ACS Nano* **2019**, *13*, 1849–1860.
[22] M. Zhang, Y. Chen, C. Wu, R. Zheng, Y. Xia, E. G. Saccuzzo, T. K. H. Trinh, E. A. Q. Mondarte, E. Nakouzi, B. Rad, B. A. Legg, W. J. Shaw, J. Tao, J. J. De Yoreo, C.-L. Chen, *Proceedings of the National Academy of Sciences* **2024**, *121*, e2412358121.
[23] Z. Zhu, H. Jin, K. Xie, S. Dai, Y. Luo, B. Qi, Z. Wang, X. Zhuang, K. Liu, B. Hu, L. Huang, J. Zhou, *Small* **2022**, *18*, 2204713.
[24] K. E. K. Sun, T. K. A. Hoang, T. N. L. Doan, Y. Yu, X. Zhu, Y. Tian, P. Chen, *ACS Appl. Mater. Interfaces* **2017**, *9*, 9681–9687.
[25] T. Wei, Y. Ren, Y. Wang, L. Mo, Z. Li, H. Zhang, L. Hu, G. Cao, *ACS Nano* **2023**, *17*, 3765–3775.
[26] R. Qin, Y. Wang, M. Zhang, Y. Wang, S. Ding, A. Song, H. Yi, L. Yang, Y. Song, Y. Cui, J. Liu, Z. Wang, S. Li, Q. Zhao, F. Pan, *Nano Energy* **2021**, *80*, 105478.
[27] S. Jin, J. Yin, X. Gao, A. Sharma, P. Chen, S. Hong, Q. Zhao, J. Zheng, Y. Deng, Y. L. Joo, L. A. Archer, *Nature Communications* **2022**, *13*, 2283.
[28] M. Yan, C. Xu, Y. Sun, H. Pan, H. Li, *Nano Energy* **2021**, *82*, 105739.
[29] X. Zhou, Y. Lu, Q. Zhang, L. Miao, K. Zhang, Z. Yan, F. Li, J. Chen, *ACS Appl. Mater. Interfaces* **2020**, *12*, 55476–55482.
[30] S. Jin, Y. Deng, P. Chen, S. Hong, R. Garcia-Mendez, A. Sharma, N. W. Utomo, Y. Shao, R. Yang, L. A. Archer, *Angewandte Chemie International Edition* **2023**, *62*, e202300823.
[31] A. Mitha, A. Z. Yazdi, M. Ahmed, P. Chen, *ChemElectroChem* **2018**, *5*, 2409–2418.

[32] F. C. Frank, J. H. Van Der Merwe, N. F. Mott, *Proceedings of the Royal Society of London. Series A. Mathematical and Physical Sciences* **1997**, *198*, 216–225.
[33] S. N. Magonov, D. H. Reneker, *Annual Review of Materials Research* **1997**, *27*, 175–222.
[34] R. Garcia, *Chem. Soc. Rev.* **2020**, *49*, 5850–5884.
[35] A. A. Sifat, J. Jahng, E. O. Potma, *Chem. Soc. Rev.* **2022**, *51*, 4208–4222.
[36] S. Akkineni, G. S. Doerk, C. Shi, B. Jin, S. Zhang, S. Habelitz, J. J. De Yoreo, *Nano Lett.* **2023**, *23*, 4290–4297.
[37] Lin Shiming, Chen Ji-Liang, Huang Long-Sun, Lin Huan-We, *Current Proteomics* **2005**, *2*, 55–81.
[38] E. Abdelrazek, A. Abdelghany, S. Badr, M. Abdelrahim, *Research Journal of Pharmaceutical, Biological and Chemical Sciences* **2016**, *7*, 1877–1890.
[39] E. M. Abdelrazek, A. M. Abdelghany, S. I. Badr, M. A. Morsi, *Journal of Materials Research and Technology* **2018**, *7*, 419–431.
[40] W.-G. Lim, X. Li, D. Reed, *Small Methods* **2024**, *8*, 2300965.
[41] T. E. Graedel, *Journal of The Electrochemical Society* **1989**, *136*, 193C.
[42] J. Hao, X. Li, S. Zhang, F. Yang, X. Zeng, S. Zhang, G. Bo, C. Wang, Z. Guo, *Advanced Functional Materials* **2020**, *30*, 2001263.
[43] R. B. Corey, R. W. Wyckoff, *Zeitschrift für Kristallographie-Crystalline Materials* **1933**, *86*, 8–18.
[44] O. K. Srivastava, E. A. Secco, *Can. J. Chem.* **1967**, *45*, 585–588.
[45] P. Geetha Devi, A. Sakthi Velu, *Journal of Materials Science: Materials in Electronics* **2016**, *27*, 10833–10840.
[46] X. Wang, L. Andrews, *J. Phys. Chem. A* **2005**, *109*, 3849–3857.
[47] I. Bear, I. Grey, I. Newnham, L. Rogers, *Aust. J. Chem.* **1987**, *40*, 539–556.
[48] I. J. Bear, I. E. Grey, I. C. Madsen, I. E. Newnham, L. J. Rogers, *Acta Crystallographica Section B* **1986**, *42*, 32–39.
[49] E. A. Secco, *Can. J. Chem.* **1988**, *66*, 329–336.
[50] R. L. Frost, Ž. Ž. Gobac, A. López, Y. Xi, R. Scholz, C. Lana, R. M. F. Lima, *Journal of Molecular Structure* **2014**, *1063*, 251–258.
[51] T. Stanimirova, R. Nikolova, N. Petrova, *Crystals* **2024**, *14*, DOI 10.3390/cryst14020183.
[52] K. W. Paul, M. J. Borda, J. D. Kubicki, D. L. Sparks, *Langmuir* **2005**, *21*, 11071–11078.
[53] A. D. Becke, *The Journal of Chemical Physics* **2014**, *140*, 18A301.
[54] M. Valiev, E. J. Bylaska, N. Govind, K. Kowalski, T. P. Straatsma, H. J. J. Van Dam, D. Wang, J. Nieplocha, E. Apra, T. L. Windus, W. A. de Jong, *Computer Physics Communications* **2010**, *181*, 1477–1489.
[55] E. Aprà, E. J. Bylaska, W. A. de Jong, N. Govind, K. Kowalski, T. P. Straatsma, M. Valiev, H. J. J. van Dam, Y. Alexeev, J. Anchell, V. Anisimov, F. W. Aquino, R. Atta-Fynn, J. Autschbach, N. P. Bauman, J. C. Becca, D. E. Bernholdt, K. Bhaskaran-Nair, S. Bogatko, P. Borowski, J. Boschen, J. Brabec, A. Bruner, E. Cauët, Y. Chen, G. N. Chuev, C. J. Cramer, J. Daily, M. J. O. Deegan, T. H. Dunning Jr., M. Dupuis, K. G. Dyall, G. I. Fann, S. A. Fischer, A. Fonari, H. Früchtl, L. Gagliardi, J. Garza, N. Gawande, S. Ghosh, K. Glaesemann, A. W. Götz, J. Hammond, V. Helms, E. D. Hermes, K. Hirao, S. Hirata, M. Jacquelin, L. Jensen, B. G. Johnson, H. Jónsson, R. A. Kendall, M. Klemm, R. Kobayashi, V. Konkov, S. Krishnamoorthy, M. Krishnan, Z. Lin, R. D. Lins, R. J. Littlefield, A. J. Logsdail, K. Lopata, W. Ma, A. V. Marenich, J. Martin del Campo, D. Mejia-Rodriguez, J. E. Moore, J. M. Mullin, T. Nakajima, D. R. Nascimento, J. A. Nichols, P. J. Nichols, J.

Nieplocha, A. Otero-de-la-Roza, B. Palmer, A. Panyala, T. Pirojsirikul, B. Peng, R. Peverati, J. Pittner, L. Pollack, R. M. Richard, P. Sadayappan, G. C. Schatz, W. A. Shelton, D. W. Silverstein, D. M. A. Smith, T. A. Soares, D. Song, M. Swart, H. L. Taylor, G. S. Thomas, V. Tipparaju, D. G. Truhlar, K. Tsemekhman, T. Van Voorhis, Á. Vázquez-Mayagoitia, P. Verma, O. Villa, A. Vishnu, K. D. Vogiatzis, D. Wang, J. H. Weare, M. J. Williamson, T. L. Windus, K. Woliński, A. T. Wong, Q. Wu, C. Yang, Q. Yu, M. Zacharias, Z. Zhang, Y. Zhao, R. J. Harrison, *The Journal of Chemical Physics* **2020**, *152*, 184102.

[56] Z. Cai, J. Wang, Z. Lu, R. Zhan, Y. Ou, L. Wang, M. Dahbi, J. Alami, J. Lu, K. Amine, Y. Sun, *Angewandte Chemie International Edition* **2022**, *61*, e202116560.

[57] P. C. T. D'Ajello, M. A. Fiori, A. A. Pasa, Z. G. Kipervaser, *Journal of The Electrochemical Society* **2000**, *147*, 4562.

[58] J. Eaves-Rathert, K. Moyer, M. Zohair, C. L. Pint, *Joule* **2020**, *4*, 1324–1336.

[59] Y. Yang, H. Yang, R. Zhu, H. Zhou, *Energy Environ. Sci.* **2023**, *16*, 2723–2731.

[60] W. Yuan, X. Nie, G. Ma, M. Liu, Y. Wang, S. Shen, N. Zhang, *Angewandte Chemie International Edition* **2023**, *62*, e202218386.

[61] J. Zhang, W. Huang, L. Li, C. Chang, K. Yang, L. Gao, X. Pu, *Advanced Materials* **2023**, *35*, 2300073.

[62] Z. Cai, J. Wang, S. Lian, J. Chen, F. Lang, Z. Li, Q. Li, *Advanced Functional Materials* **2024**, *n/a*, 2401367.

[63] M. Zhang, P. A. Salvador, G. S. Rohrer, *Journal of the European Ceramic Society* **2021**, *41*, 319–325.

# Achieving electrode smoothing by controlling the nucleation phase of metal deposition through polymer-substrate binding

Ying Xia, Duo Song, Mingyi Zhang, Zheming Wang, Chenyang Shi, Jingshan S. Du, Sun Hae Ra Shin, Mark H. Engelhard, Praveen K. Thallapally, Christine A. Orme, Jinhui Tao, Maria L. Sushko,* James. J. De Yoreo,* and Jun Liu*

Y. Xia, J. J. De Yoreo, J. Liu
Department of Materials Science and Engineering
University of Washington
Seattle, WA 98115, USA
E-mail: jliuuw1@uw.edu

Y. Xia, D. Song, M. Zhang, Z. Wang, C. Shi, J. S. Du, J. Tao, M. L. Sushko, J. J. De Yoreo
Physical and Computational Sciences Directorate
Pacific Northwest National Laboratory
Richland, WA 99354, USA
E-mail: James.DeYoreo@pnnl.gov
E-mail: Maria.Sushko@pnnl.gov

S. H. R. Shin, M. H. Engelhard, P. K. Thallapally, J. Liu
Energy and Environment Directorate
Pacific Northwest National Laboratory
Richland, WA 99354, USA

C. A. Orme
Lawrence Livermore National Laboratory
7000 East Avenue, Livermore, CA 94550, USA

Corresponding author: Maria.Sushko@pnnl.gov, James.DeYoreo@pnnl.gov, jliuuw1@uw.edu

**This PDF file includes**:
Materials and Methods
Supplementary Text
Fig. S1 to S26
Caption for Video S1 and Video S2
References
**Other supplementary material for this manuscript includes the following:**
Video S1 and Video S2

# Section 1. Materials and Experimental Methods

**Electrolytes Preparation**

All cell components and the electrolyte reservoir were sonicated in ultrapure water (Milli-Q, 18.2 MΩ·cm) for 10 minutes and dried using an $N_2$ gun. Electrolytes were prepared using ultrapure water (Water OmniTrace Ultra, Sigma Aldrich), zinc sulfate heptahydrate (99.5%, for analysis, Thermo Scientific Chemicals), and polyethylene oxide (PEO, M.W. 100,000, Thermo Scientific Chemicals). Solutions with varying PEO concentrations were obtained by diluting stock PEO solutions.

The 5 wt% PEO stock solution was prepared by dissolving PEO in water through repeated cycles of sonication and vortexing until fully dissolved. The solution was then centrifuged at 13,400 rpm for 2 minutes to separate any insoluble components. The clear supernatant was collected and used as the stock solution. For experiments requiring deoxygenation, the electrolytes were purged with $N_2$ gas for at least 10 minutes to remove dissolved oxygen.

**Cu and Zn Substrates Preparation**

Polycrystalline Copper foil (9 μm thick, Purity≥ 99.8%, MTI Corporation), punched into 12.7 mm-diameter round plates, was used as the working electrode. Polycrystalline Zinc foil (0.25 mm thick, 99.98% metals basis, Thermo Scientific Chemicals), cut into 10 mm × 10 mm pieces, served as AFM substrates for studying zinc oxidation reactions and PEO adsorption behaviors.

To remove carbon contamination and surface oxides, Zn foils were first rinsed multiple times with acetone and Milli-Q water, followed by a 90-second treatment in 0.1 M $H_2SO_4$. Cu foils underwent a similar cleaning protocol but without the initial acetone rinse. Ethanol was then used to remove residual acid, and the foils were dried with an $N_2$ gun. Before assembly into the AFM cell, the treated Cu and Zn foils underwent a 3-minute plasma treatment using an $N_2$ source. Cu was cycled until reaching the steady state to produce the electrode surface, which forms the Cu-Zn interfacial alloy based on the need.[1]

***In-situ* Electrochemical Atomic Force Microscopy (EC-AFM) Measurements**

EC-AFM experiments were performed using the Multimode Nanoscope 8 AFM (Bruker) equipped with a standard liquid AFM cell (Bruker). Electrochemical conditions were controlled by coupling the AFM system with a 600E potentiostat (CH Instruments). A homemade counter electrode, fabricated from Zn wires (1.0 mm diameter, 99.95% metals basis, Thermo Scientific Chemicals), was inserted into the testing solution via a dedicated channel in the liquid AFM cell and stabilized using a custom-design sealing cap. The electrolyte solution consists of 0.1 M $ZnSO_4$ either with or without the addition of 0.1 wt% PEO.

The working electrode, a round Cu foil plate after special treatments following above protocols (see **Cu and Zn Substrates Preparation**) mounted on the AFM sample disk, was connected to the 600E potentiostat using a nickel wire. The electrochemical signal was applied in a two-electrode configuration controlled by the 600E potentiostat. Before electrodeposition, the open circuit potential (OCP) was measured and remained stable at 1.05 ± 0.03 V. During electrodeposition, either a constant voltage of -0.015 V was applied, or the voltage was gradually decreased from 0 V to -0.015 V at a uniform scan rate of 0.01 mV/s over 25 minutes.

All *in-situ* EC-AFM imaging was performed in liquid using either PeakForce Tapping™ mode or amplitude modulation mode at room temperature (23 °C) on the Nanoscope 8 AFM (J scanner, Bruker). An FM-20 AFM probe from Nanoworld, with a spring constant k = 2.8 N/m, and a resonance frequency f=75 kHz, was used for imaging. The probe was plasma-cleaned for 3 minutes before imaging. The typical scanning rate is 1 Hz, and the drive frequency was set to approximately 75 kHz. The amplitude setpoint was controlled to maintain image quality without changing the surface topography. Offline data processing and analysis were performed using the Nanoscope Analysis 2.0 software package (Bruker).

The slope distribution of the AFM images was analyzed using Gwyddion's Slope Distribution function. This function calculates the horizontal and vertical derivatives by fitting a local plane through the neighborhood of each point and determining its gradient, which defines the horizontal and vertical coordinates in the resulting data field.

***In-situ* AFM Measurements for Zn Oxidation and PEO Adsorption on Cu/Zn**

All *in-situ* AFM measurements capturing morphology and phase evolution during the Zn oxidation and PEO adsorption processes were performed in liquid using amplitude modulation mode on a Cypher VRS AFM (Asylum Research). Imaging was conducted with either Arrow™ UHF AuD probes (NanoWorld, k = 1 N/m - 5.0 N/m, f = 2000 kHz) or SNL-10 probes (Bruker, k = 0.24 N/m, tip radius = 2 nm, f = 56 kHz). The probe was plasma-cleaned for 3 minutes before imaging. The typical scanning rate has a range of 1-13 Hz, with the drive frequency set at approximately 1000 kHz for the Arrow™ UHF probes and around 45 kHz for the SNL-10 probes. The amplitude setpoint was controlled to maintain image quality without changing the surface topography. Offline data processing and analysis were performed using Gwyddion SPM data analysis software.

**Adhesion Force Mapping**

Adhesion force mapping was performed in liquid using Contact Fast Force Mapping mode on a Cypher VRS AFM (Asylum Research) with Arrow™ UHF AuD probes (NanoWorld), featuring a spring constant k ranging from 1 N/m to 5.0 N/m and a resonance frequency f = 2000 kHz. Before the experiment, the probes were cleaned by UV-Ozone for 5-10 mins. The typical scanning rate is 300 kHz, with 256 × 256 force curves collected per image. A deflection setpoint of approximately 0.1 V was used and carefully controlled to ensure high-quality force curves while preserving the surface topography. Image processing and analysis were carried out using the Igor Pro software package.

**AFM under Flowing Conditions**

Solution exchange was facilitated by PTFE tubing connected to a syringe pump, with a flow rate of 0.1 mL/min. To ensure complete replacement, at least 1 mL of new solution was introduced between each imaging condition.

**Sample Preparation for Photo-induced Force Microscopy (PiFM) and Water Contact Angle Measurements**

Cu foils were first plasma-treated for 3 minutes. Then 100 μL of PEO solution at varying concentrations (0.001 wt%, 0.1 wt%) was drop-cast onto the plasma-cleaned Cu surface and incubated at room temperature in a sealed chamber with a relative humidity of ~100% for 2 hours. After incubation, the solutions on the Cu surface were gently exchanged with Milli-Q water 10 times to remove unbound PEO molecules. The wet PEO-coated Cu samples were then allowed to dry slowly in ambient air.

**PiFM Measurement**

All samples were analyzed in the air with a VistaScope PiFM (Molecular Vista Inc.), integrated with a Laser Tune QCL featuring a wavenumber resolution of 0.5 $cm^{-1}$ and a tuning range from 800 to 1800 $cm^{-1}$. The microscope operated in dynamic mode, utilizing HQ/Cr-Au probes (MikroMasch). Data processing was performed using Surface Works software (Molecular Vista Inc.).

**Water Contact Angle Measurement**

Sample preparation followed the same protocol as for PiFM testing. Water contact angle measurements were performed using the FTÅ 200 drop shape analysis system. Each sample was tested 3–5 times, and the final water contact angle was determined by averaging the multiple measurements.

**Transmission Electron Microscopy (TEM) Measurement**

After electrodeposition, the electrolytes in the AFM liquid cell were immediately replaced with Milli-Q water to rinse the system. The freshly deposited Zn on Cu substrates was then sonicated in ethanol to exfoliate the Zn from the Cu surface. Multiple droplets of the resulting Zn-ethanol solution were carefully pipetted onto a TEM grid and allowed to dry under ambient conditions. TEM was performed on an FEI Titan Environmental TEM [extreme-brightness Schottky field-emission gun (X-FEG), 300 kV] equipped with a CEOS double-hexapole aberration corrector (CETCOR) for the image-forming lenses and a Gatan UltraScan 1000 charge-coupled device (CCD) scintillation camera.

**Grazing Incidence X-ray Diffraction (GIXRD)**

XRD data were collected using a Rigaku SmartLab XE diffractometer equipped with a Cu rotating anode ($\lambda$ = 1.5418 Å) and an incident angle of 1°.

**Scanning Electron Microscopy (SEM) and Electron backscatter diffraction (EBSD)**

SEM images were acquired using an Apreo Variable Pressure SEM (ThermoFisher) equipped with an EBSD detector (Oxford). The SEM imaging was performed with a 10 kV beam and a current intensity of 3.2 nA using the ETD detector. EBSD patterns were recorded with a step size of 0.1 µm and an acceleration voltage of 20 kV. The patterns were indexed using the AZtecCrystal Processing Software, with acquisition parameters for copper defined as: a = 3.61 Å, b = 3.61 Å, c = 3.61 Å, α = 90.00°, β = 90.00°, γ = 90.00° (space group 225).

**Attenuated Total Reflectance Fourier Transform Infrared Spectroscopy (ATR-FTIR)**

ATR-FTIR measurements were done using a LUMOS infrared microscope in attenuated total reflection (ATR) mode (LUMOS, Bruker Corp.) and acquired using the OPUS 8.2.28 software. In the ATR-FTIR measurement, an ATR-FTIR measurement in air prior to the introduction of water or 0.1 M $ZnSO_4$ solution was used as the ATR-FTIR background for each time series of measurements and the time between the liquid introduction and the first measurement was minimized (1 to 2 minutes). Each of the spectra are the average of 32 scans. A linear background has been subtracted from the raw spectra.

**X-ray Photoelectron Spectroscopy (XPS)**

XPS spectra were measured using a Thermo Fisher NEXSA instrument with a monochromatic Al Kα (1468.7 eV) excitation source and double-focusing hemispherical analyzer with multielement input lens and a detector with 128 channels. The indecent X-ray beam was normal to the sample, while the photoelectron detector was positioned 60° off normal from the sample. The pass energy was 50 eV, the step size was 0.1 eV, and the dwell time was 50 ms for the high-resolution spectra. The full-width-at-half-maximum (FWHM) was measured to be 0.82 eV for the Cu $2p_{3/2}$ peak using the same conditions as those at which the narrow scan spectra were collected.

## Section 2. Computational Methods

*Density Functional Theory (DFT)*

All DFT[2] geometry optimizations in this study were performed with the pseudopotential plane-wave NWPW module[3] implemented in the NWChem software package[4,5]. The DFT Local Density Approximation (LDA) functional[6] was used throughout to account for the exchange-correlation energy. In our plane-wave calculations, the valence electron interactions with the atomic core were approximated with generalized norm-conserving Hamann pseudopotentials[7] for C, O, and H. These pseudopotentials were constructed using the following core radii: $r_{cs}$ = 0.8 a.u., and $r_{cp}$ = $r_{cd}$ = 0.85 a.u. for C; $r_{cs}$ = $r_{cp}$ = $r_{cd}$ = 0.7 a.u. for O; and $r_{cs}$ = $r_{cp}$ = 0.8 a.u. for H. Norm-conserving Troullier-Martins pseudopotentials[8] were applied for Cu and Zn. The following core radii were used to generate these pseudopotentials: $r_{cs}$ = $r_{cd}$ = 2.055 a.u., and $r_{cp}$ = 2.299 a.u. for Cu and $r_{cs}$ = 1.870 a.u., $r_{cp}$ = 1.850 a.u., and $r_{cd}$ = 1.600 a.u. for Zn. All the pseudopotentials were modified to the separable form suggested by Kleinman and Bylander.[9] Restricted calculations were performed since all the systems are closed shell. The electronic wavefunctions were expanded using a plane-wave basis with periodic boundary conditions at the Γ-point (1×1×1 k-point mesh) with a wavefunction cutoff energy of 40 Ry and a density cutoff energy of 80 Ry. The Cu slab was frozen during geometry optimization for all the diethylene glycol (DEG)- and H2O- saturated systems. Other atomic positions were relaxed using conjugate-gradient algorithm until the forces on the atoms were converged to less than $5\times10^{-3}$ Hartree/Bohr and the total energy was converged to less than $10^{-3}$ Hartree.

*Classical DFT*

Within cDFT, solvent density is optimized along with the densities of all other species and as such solvent structuring and solvation effects are intrinsically included in all components of the free energy.[10,11] The total free energy functional encompasses the contributions from Coulomb interactions ($F_C^{ex}$), electrostatic correlations ($F_{el}^{ex}$), hard sphere repulsion ($F_{hs}^{ex}$), short-range interactions ($F_{sh}^{ex}$) between the species, image interactions ($F_{im}^{ex}$) and ion-surface interactions ($F_{ion-s}^{ex}$):

$$F^{ex} = F_C^{ex} + F_{el}^{ex} + F_{hs}^{ex} + F_{sh}^{ex} + F_{im}^{ex} + F_{ion-s}^{ex}$$

All contributions to the excess free energy, except the short-range term, are calculated analytically from first principles using the Mean Spherical Approximation, the Fundamental Measure Theory and Lifshitz theory of dispersion interactions (full details of the method are provided below). The short-range interactions between all mobile species are given by

$$F_{sh}^{ex} = \frac{1}{2}\int_\Omega\int_\Omega dr\,dr'\,\Sigma\rho_\alpha(r)\rho_\beta(r')\Phi_{\alpha\beta}(|r-r'|)$$

where $\Phi_{\alpha\beta}$ is the square-well potential of depth equal to the equilibrium pair-wise interaction energies calculated using plane-wave DFT. These interactions include ion-solvent interactions (solvation) and ion-surface chemical interactions. Density profiles of solution species are calculated within cDFT via the minimization of the free energy functional with respect to the densities of all the species. The densities satisfy the following equation

$$\rho_i(r) = \rho_i^{bulk}(r)exp\left(-\frac{q_i\varphi(r)}{kT} - \frac{1}{kT}\frac{\delta\left(F_{el}^{ex} + F_{hs}^{ex} + F_{sh}^{ex} + F_{ion\_s}^{ex}\right)}{\delta\rho_i(r)}\right)$$

where $\rho_i^{bulk}$ are the bulk values for the densities of solution species, such as ions and water molecules. Poisson's equation is solved to calculate the electrostatic potential ($\varphi(r)$) and the corresponding term in Eq. (3) accounts for Coulomb and image interactions. The resulting system of equations is solved iteratively to self-consistency using the relaxed Gummel iterative procedure.[11]

Chemical potential calculations utilize the notion that although the total chemical potential for each of the species is constant, individual components of chemical potential are position dependent, vary with local concentration and reflect the nature of the driving forces for nucleation

and growth:

$$\mu^{bulk} = \mu^{id}(r) + \mu^{ex}(r),$$

where $\mu^{id}(r)$ is the ideal chemical potential and the excess chemical potential is calculated as

$$\mu^{ex}(r) = \mu_C^{ex}(r) + \mu_{el}^{ex}(r) + \mu_{hs}^{ex}(r) + \mu_{sh}^{ex}(r) + \mu_{im}^{ex}(r) + \mu_{ion-s}^{ex}(r)$$

where the subscripts have the same meanings as above.

Free energy functional. To determine the equilibrium water and ion distributions via cDFT, the total Helmholtz free energy functional is minimized with respect to the densities of all the species in the presence of rigid nanoparticles or mineral surfaces. For this optimization, it is convenient to partition the total free energy of the system into so-called ideal (F$_{id}$) and excess components (F$_{ex}$).[12] The ideal free energy corresponds to the non-interacting system and is determined by the configurational entropy contributions from water and small ions,

$$F^{id} = kT \sum_i^N \int_\Omega (\rho_i(r) \log \rho_i(r) - \rho_i(r)) dr$$

where $k$ is Boltzmann's constant, $T$ is the temperature, $\rho_i$ is the density profile of ion and water species $i$, $N$ is the number of species, $r \in \Omega$ is the ion coordinate, and $\Omega$ is the calculation domain. The excess free energy is generally not known exactly but can be approximated by

$$F^{ex} = F_{EDL}^{ex} + F_{hs}^{ex} + F_{el}^{ex} + F_{hydr}^{ex} + F_{ion\_s}^{ex}$$

where $F_{EDL}^{ex} = F_C^{ex} + F_{im}^{ex}$ describes first-order electrostatics and includes direct Coulomb term and image terms, $F_{hs}^{ex}$ is the hard sphere repulsion term, $F_{el}^{ex}$ is the electrostatic ion correlation term, $F_{hydr}^{ex}$ is the ion hydration term, and $F_{ion\_s}^{ex}$ describes ion-surface short-range interactions.

Poisson equation for first-order electrostatics. The electric double layer contribution to the free energy ($F_{EDL}^{ex}$) includes direct Coulomb and image interactions and is evaluated through the solution of Poisson's equation

$$-\nabla \cdot \varepsilon(r) \nabla \varphi(r) = \rho_f(r) + \sum_i q_i \rho_i(r)$$

for the electrostatic potential, $\varphi(r)$, where $\rho_f(r)$ is the fixed charge density on crystal faces, $\rho_i(r)$ is the density of mobile species with charge $q_i$, $\varepsilon(r)$ is the dielectric coefficient equal to 78.5 in solution and to static dielectric constant of the mineral. The discrete distribution of charges on mineral faces is constructed on a 2D grid using trilinear interpolation.

The corresponding EDL contribution to the free energy is calculated as

$$F_{EDL}^{ex} = \sum_i \int_\Omega q_i \rho_i(r) \varphi(r) dr$$

which provides an exact solution for first-order electrostatics.

Fundamental Measure Theory of excluded volume effects. Hard sphere repulsive interactions describe ion and water many-body interactions in condensed phase due to density fluctuations. These interactions were described using a Fundamental Measure theory (FMT).[13] The approach is based on the solution of the Ornstein-Zernike equation for direct correlation function using the Percus–Yevick approximation and yields the following form of the corresponding component of the free energy:[14]

$$F_{hs}^{ex} = kT \int \Phi_{hs}[n_\omega(\mathbf{r})]d\mathbf{r}$$

where the hard-sphere free energy density $\Phi_{hs}$ is a functional of four scalar and two vector weighted densities ($n_\omega(\mathbf{r})$) and has the form

$$\Phi_{hs}(r) = -n_0 \ln(1-n_3) + \frac{n_1 n_2}{1-n_3} + \left[\frac{1}{36\pi n_3^2}\ln(1-n_3) + \frac{1}{36\pi n_3(1-n_3)^2}\right]n_2^3$$

$$-\frac{\mathbf{n}_1 \cdot \mathbf{n}_2}{1-n_3} - \left[\frac{1}{12\pi n_3^2}\ln(1-n_3) + \frac{1}{12\pi n_3(1-n_3)^2}\right]n_2(\mathbf{n}_2 \cdot \mathbf{n}_2)$$

where scalar ($\alpha$ = 0, 1, 2, 3) and vector ($\beta$ = 1, 2) weighted densities are defined as

$$n_\alpha(\mathbf{r}) = \sum_i \int_\Omega \rho_i(\mathbf{r}')\omega_i^{(\alpha)}(\mathbf{r}'-\mathbf{r})d\mathbf{r}'$$

$$\mathbf{n}_\beta(\mathbf{r}) = \sum_i \int_\Omega \rho_i(\mathbf{r}')\boldsymbol{\omega}_i^{(\beta)}(\mathbf{r}'-\mathbf{r})d\mathbf{r}'$$

The "weight functions" $\omega^{(\alpha)}$ and $\omega^{(\beta)}$ characterizing the geometry of particles (hard sphere with radius $R_i$ for ion species $i$) are given by:

$\omega_i^{(3)}(\mathbf{r}) = \theta(|\mathbf{r}|-R_i)$

$\omega_i^{(2)}(\mathbf{r}) = |\nabla\theta(|\mathbf{r}|-R_i)| = \delta(|\mathbf{r}|-R_i)$

$\boldsymbol{\omega}_i^{(2)}(\mathbf{r}) = \nabla\theta(|\mathbf{r}|-R_i) = \frac{\mathbf{r}}{r}\delta(|\mathbf{r}|-R_i)$

$\omega_i^{(0)}(\mathbf{r}) = \omega_i^{(2)}(\mathbf{r})/(4\pi R_i^2)$

$\omega_i^{(1)}(\mathbf{r}) = \omega_i^{(2)}(\mathbf{r})/(4\pi R_i)$

$\boldsymbol{\omega}_i^{(1)}(\mathbf{r}) = \boldsymbol{\omega}_i^{(2)}(\mathbf{r})/(4\pi R_i)$

In the preceding formulae, $\theta$ is the Heaviside step function with $\theta(x) = 0$ for $x > 0$ and $\theta(x) = 1$ for $x \le 0$, and $\delta$ denotes the Dirac delta function.

Mean Spherical Approximation of ion-ion electrostatic correlations. To treat correlations resulting from electrostatic interactions between charged species on the same footing as those resulting from hard sphere excluded volume interactions, the Mean Spherical Approximation[15,16] was employed to solve the Ornstein-Zernike equation with respect to the electrostatic direct correlation function. Taylor expansion of the electrostatic free energy was cut after the second order. Then the electrostatic correlation component of the free energy ($F_{el}^{ex}$) is

$$F_{el}^{ex} = F_{el}^{ex}[\{\rho_i^{bulk}\}] - kT \int d\mathbf{r} \sum_{i=+,-} \Delta C_i^{(1)el}(\rho_i(\mathbf{r}) - \rho_i^{bulk}) -$$

$$\frac{kT}{2} \iint d\mathbf{r}d\mathbf{r}' \sum_{i,j=+,-} \Delta C_{ij}^{(2)el}(|\mathbf{r}-\mathbf{r}'|)(\rho_i(\mathbf{r}) - \rho_i^{bulk})(\rho_j(\mathbf{r}') - \rho_j^{bulk})$$

where $\rho_i^{bulk}$ are the bulk densities of charged species and the first and second-order direct correlation functions are defined as

$$\Delta C_i^{(1)el} = -\mu_i^{el}/kT$$

$$\Delta C_{ij}^{(2)el}(|\mathbf{r}-\mathbf{r}'|) = \begin{cases} -\frac{q_i q_j e^2}{kT\varepsilon}\left[\frac{2B}{\sigma_{ij}} - \left(\frac{2B}{\sigma_{ij}}\right)^2(|\mathbf{r}-\mathbf{r}'|) - \frac{1}{(|\mathbf{r}-\mathbf{r}'|)}\right], & (|\mathbf{r}-\mathbf{r}'|) \leq \sigma_{ij} \\ 0, & (|\mathbf{r}-\mathbf{r}'|) > \sigma_{ij} \end{cases}$$

with

$$B = [\xi + 1 - (1 + 2\xi)^{1/2}]/\xi$$

$$\xi^2 = \kappa^2 \sigma_{ij}^2 = \left[\frac{e^2}{\varepsilon kT} \sum_i q_i^2 \rho_i^{bulk}\right] \sigma_{ij}^2$$

In the above equations, $\mu_i^{el}$ is the chemical potential of the mobile ions, $\kappa$ is the inverse Debye length and the contact distance $\sigma_{ij} = (\sigma_i + \sigma_j)/2$.

In many cases these interactions lead to the overall attractive interactions between like-charged surfaces[17–21] or polyelectrolytes[22–24] even in low salt conditions[25].

Ion hydration interactions. The short-range attractive hydration interactions between ions (denoted as "ion") and water "molecules" (denoted as "w") in electrolyte solution are given by

$$F_{sh}^{ex} = \frac{1}{2}\int_\Omega \int_\Omega d\mathbf{r}d\mathbf{r}' \sum_{\alpha,\beta=ion,w} \rho_\alpha(\mathbf{r})\rho_\beta(\mathbf{r}')\Phi_{\alpha\beta}(|\mathbf{r}-\mathbf{r}'|)$$

where $\Phi_{\alpha\beta}(|\mathbf{r}-\mathbf{r}'|)$ is the square-well potential

$$\Phi_{\alpha\beta}(|\boldsymbol{r}-\boldsymbol{r}'|) = \begin{cases} \infty, & |\boldsymbol{r}-\boldsymbol{r}'| < \sigma_{\alpha\beta} \\ -\tau, & \sigma_{\alpha\beta} \leq |\boldsymbol{r}-\boldsymbol{r}'| < 1.2\sigma_{\alpha\beta} \\ 0, & |\boldsymbol{r}-\boldsymbol{r}'| \geq 1.2\sigma_{\alpha\beta} \end{cases}$$

with $\sigma_{\alpha\beta}$ equal to the contact distance between species α and β and depth $\tau$ equal to the scaled hydration enthalpy of the ions.

Ion-surface dispersion interactions. The ion-surface dispersion interactions are calculated using Lifshitz theory and are given by

$$F_{ion-s}^{ex} = -\frac{\hbar}{(4\pi)^2\varepsilon_0 d^3}\sum_i \int_\Omega \rho_i(\boldsymbol{r})d\boldsymbol{r}\int_0^\infty d\omega \frac{\alpha_i^*(i\omega)}{\varepsilon(i\omega)}\frac{\varepsilon(i\omega)-\varepsilon_{surf}(i\omega)}{\varepsilon(i\omega)+\varepsilon_{surf}(i\omega)} = \sum_i \int_\Omega \rho_i(\boldsymbol{r})d\boldsymbol{r}\frac{B_i}{d^3}$$

where $\alpha_i^*(i\omega)$ is the frequency-dependent excess polarizability of an ion of species $i$, i.e., the difference in polarizability between the solvated ion and pure solvent and $\varepsilon(i\omega)$, $\varepsilon_{surf}(i\omega)$ are the dielectric functions of the solvent and the surface, respectively. $d$ is the distance between the ion and the surface. Note that $d \geq \sigma_i/2$, as dictated by excluded volume interactions.

In this work, we generated the Zn clusters from the hexagonal bulk crystal structure of Zn with lattice parameters of a = 2.61 Å, b = 2.61 Å, c = 4.87 Å, α = 90°, β = 90°, and γ = 120°. For the Cu slab, the lattice parameters of the cubic unit cell are a = b = c = 3.58 Å.

## Section 3. Figure S1 to S26 and Table S1

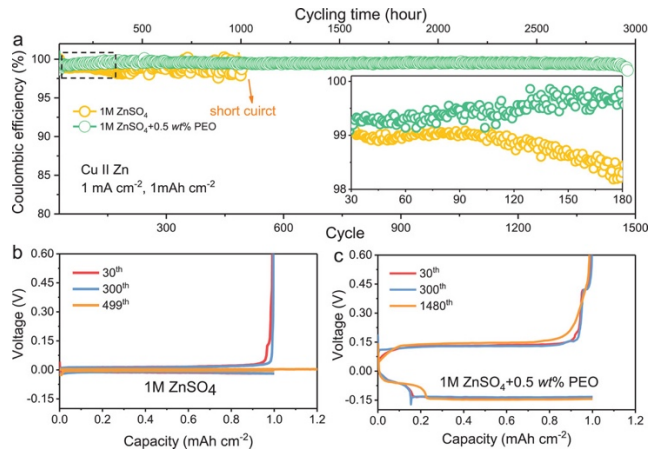

**Figure S1. Electrochemical performance of Zn metal anodes.** (a) Coulombic efficiency and cycling performance of Zn metals in Cu || Zn cells under current density of 1 mA cm$^{-2}$ for 1 mAh cm$^{-2}$. (b, c) voltage profiles of Zn stripping/plating during cycling in 1 M ZnSO$_4$ and 1 M ZnSO$_4$ + 0.5 wt.% PEO electrolytes, respectively.[26]

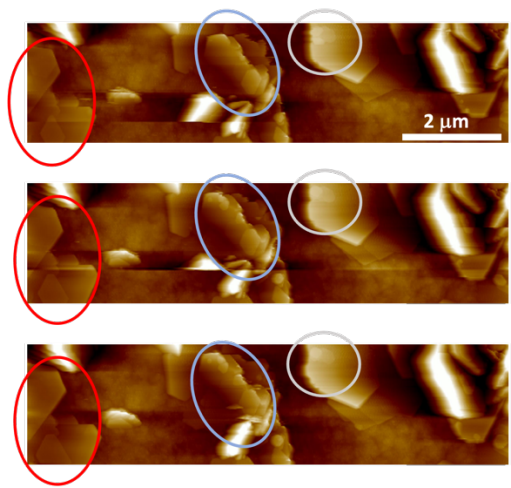

**Figure S2**. *In-situ* **EC-AFM images show the growth of Zn plates.** Colored circles highlight layer-by-layer growth.

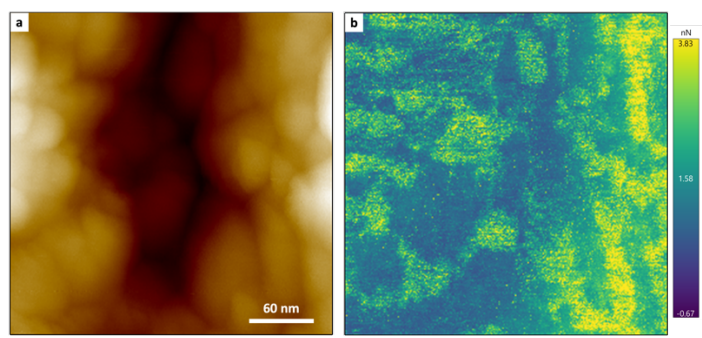

**Figure S3**. **PEO adsorption on Cu substrate in 0.1 wt% PEO solution.** (a) The AFM height image of the selected position on the Cu substrate. (b) Adhesion force map of the selected position shown in (a) in 0.1 wt% PEO solution obtained by CFFM mode (color range from -0.67 to 3.83 nN).

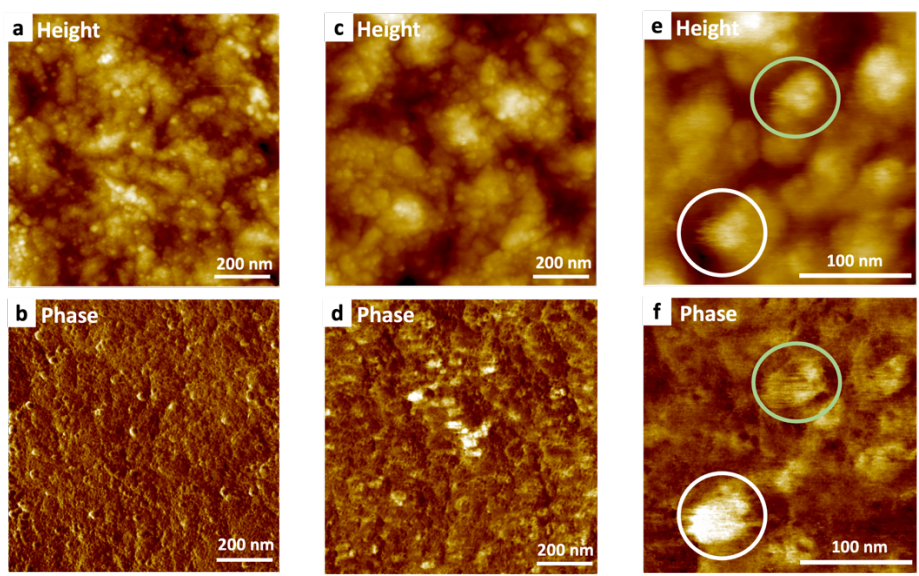

**Figure S4. Correlated height and phase images of Cu substrates in pure water and 0.1 wt% PEO solution.** (a, b) The height (a) and phase (b) images of Cu substrate in water. (c, d) The height (c) and phase (d) images of Cu substrate in 0.1 wt% PEO solution. (e, f) The zoomed-in height (e) and phase (f) image of Cu substrate in 0.1wt.% PEO solution.

When soaked in pure water, the phase images obtained by *in-situ* scanning lasting about one hour (**Fig. S4a**) show that the surface keeps only one phase, which is Cu. After soaking in 0.1 wt% PEO, the new phase emerges, which is reflected by big contrasts in **Fig. S4d**. For AFM phase images, the more positive phase represents the region possessing an attractive force with the AFM probes. Conversely, a region with a more negative phase reflects a repulsive force between the sample and AFM probes. For the color contrast, the lighter color represents a more positive phase. Thus, the lighter part in the phase image represents a softer region attracting probes, and the harder sample will show a darker color. The newly formed lighter phase is softer than the surrounding regions, which might be adsorbed soft PEO polymer on hard Cu metal substrates. We then zoomed in to obtain the high-resolution images of Cu after soaking in the PEO solution (**Fig. S4e, S4f**). The lighter region in the phase image shows an amorphous polymer aggregation morphology, as highlighted in the white and green circles in **Fig. S4e, S4f**. By correlating height and phase images, it is proved that PEO adsorbed on the Cu surface with amorphous structures.

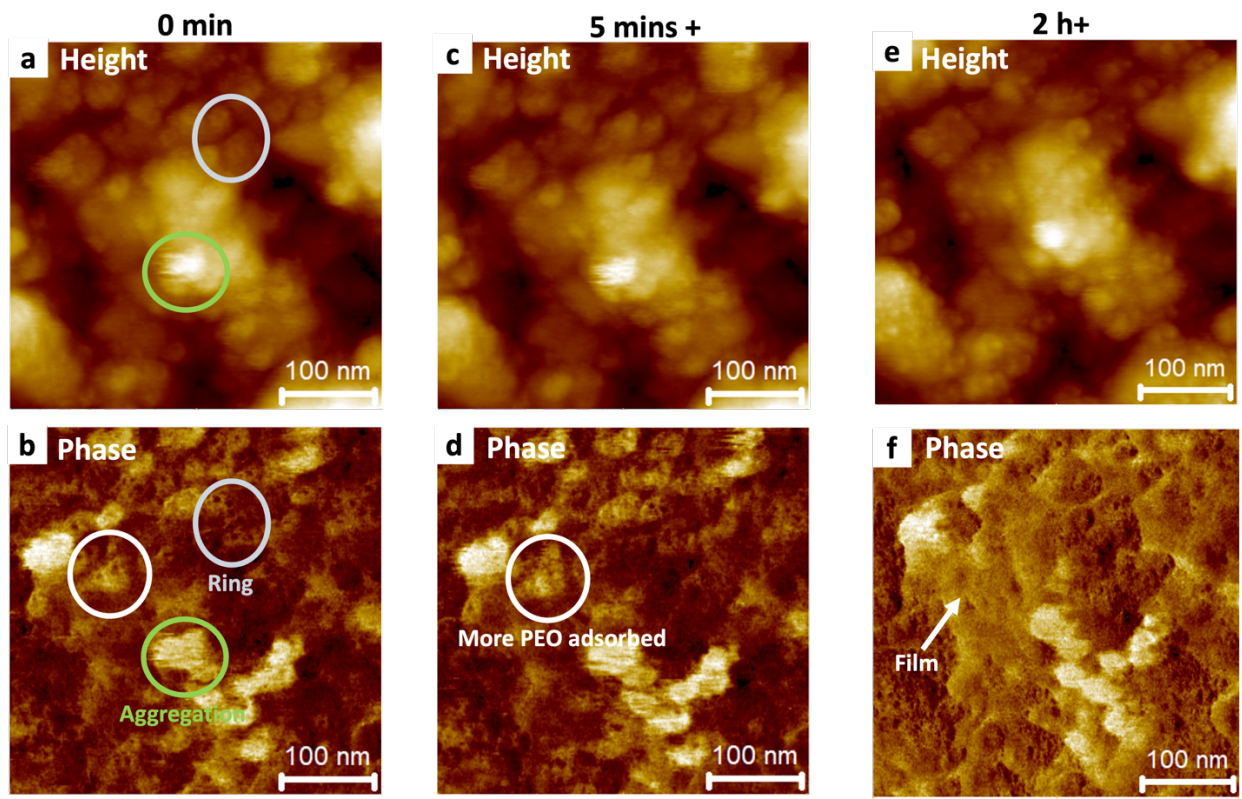

**Figure S5**. *In-situ* **AFM images of PEO adsorption on Cu in 0.1 wt% PEO solution with time show the adsorption and relaxing process.** (a, b) The height (a) and correlated phase (b) images at the early stage of PEO adsorption. (c, d) The height (c) and correlated phase (d) images captured 5 mins later than (a, b). (e, f) The height (e) and correlated phase (f) images of Cu substrates after soaking in PEO solution for two hours.

The selected *in-situ* AFM images of PEO adsorption on Cu with time are shown in **Fig. S5**. The amorphous structures of PEO include the aggregation (highlighted in the green circle in **Fig. S5a, S5b**), the ring-shaped structure (highlighted in the blue circle in **Fig. S5a, S5b**), and the porous film structure (highlighted by the white arrow in **Fig. S5f**). The adsorbed PEO seems to cover larger and larger areas of Cu substrates in the early soaking stage with time going by, as highlighted in the white circles in **Fig. S5b, S5d**. After soaking in PEO solution for 2 hours, the aggregation and the ring-shaped structure tend to relax into porous film structures with time (**Fig. S5f**). The aggregation of PEO on the Cu substrate might lead to the lower nucleation density of Zn ECD, which was observed in **Fig. 1b** compared to **Fig. 1a**. Those thick aggregations might block partial conductive Cu surfaces and hinder the Zn ECD process from happening. The adsorbed PEO with ring-shaped or porous film structures might tune the interfacial energy of Cu-electrolyte interfaces without blocking the active Cu surfaces for electron and charge transfer.[27]

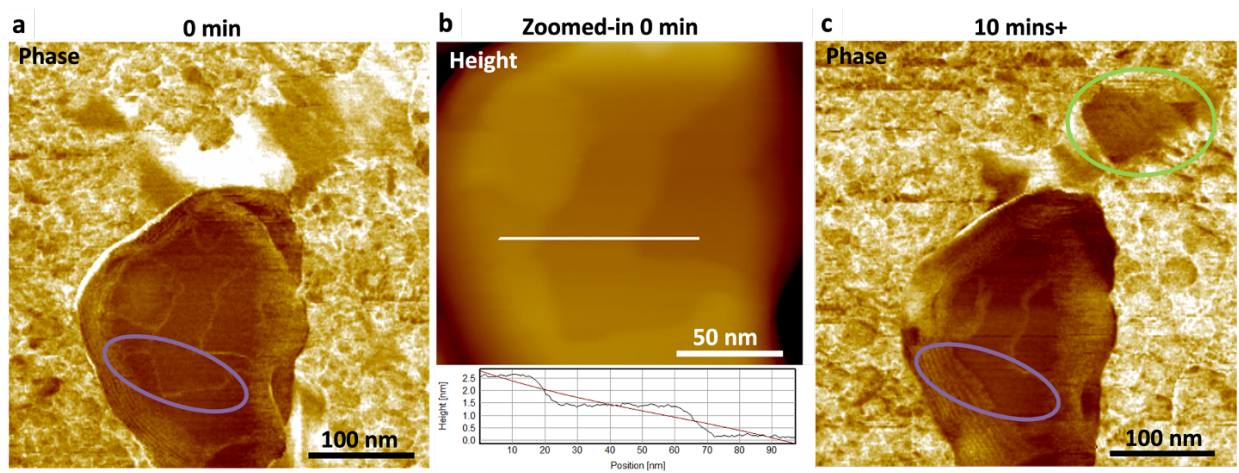

**Figure S6**. **Etching and surface reorganization of Zn on Cu in 0.1 M ZnSO₄ electrolytes without reducing voltage.** (a) The phase image of etching Zn metal on Cu substrates, (b) The zoomed-in height image of layered structure in (a). (c) The phase image captured 10 mins after (a).

There are two phases in the phase image of Zn on the Cu system shown in **Fig. S6a**: the Zn plates are darker, and the surrounding Cu substrates are brighter. The Zn metal etches into layered structures with a 1.2 nm step height, which is compatible with the height of a unit cell of Zn metal (**Fig. S6b**).[28,29] The edges of the Zn layers will move backward with time, as highlighted in purple circles in **Fig. S6a, S6c**. Except for the etching process, we also noticed newly formed structures near the etching Zn plates with a similar color contrast with Zn plates highlighted in the green circle in **Fig. S6c**, which might be the reorganization of the dissolved Zn metal or the formation of a new phase.

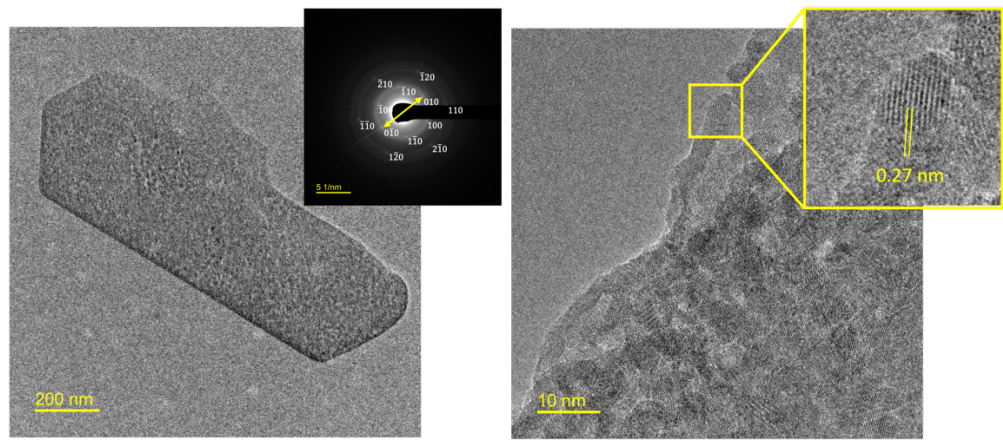

**Figure S7**. Lattice-resolved TEM images and diffraction pattern of plates show that initially formed Zn platelets oxidize to $Zn(OH)_2$ in water.

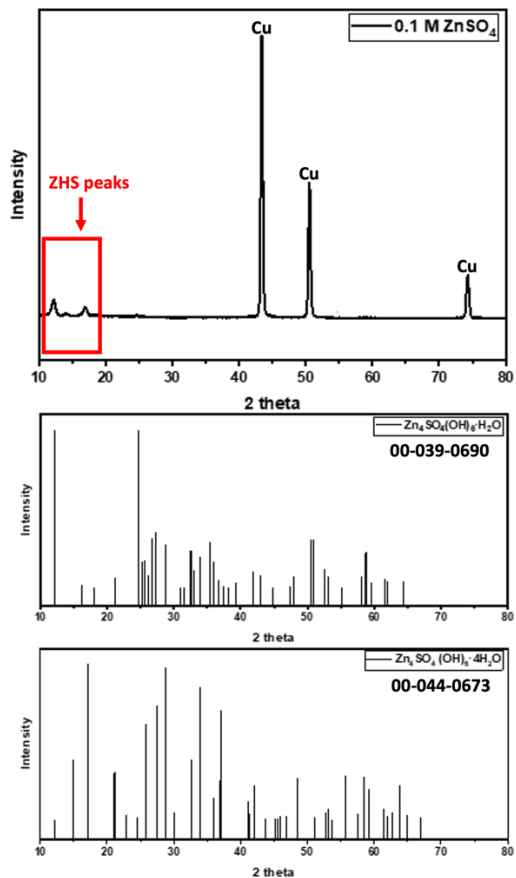

**Figure S8**. GIXRD results show that the deposited Zn crystals on Cu become $Zn_4SO_4(OH)_6 \cdot xH_2O$ (ZHS) on Cu after drying.

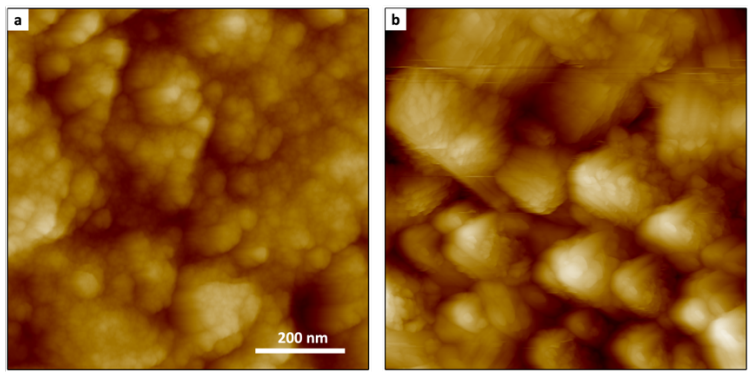

**Figure S9. Surface reorganization of Zn foil in water.** The height images of Zn metal after soaking in water for 30 minutes (a) and 115 minutes (b).

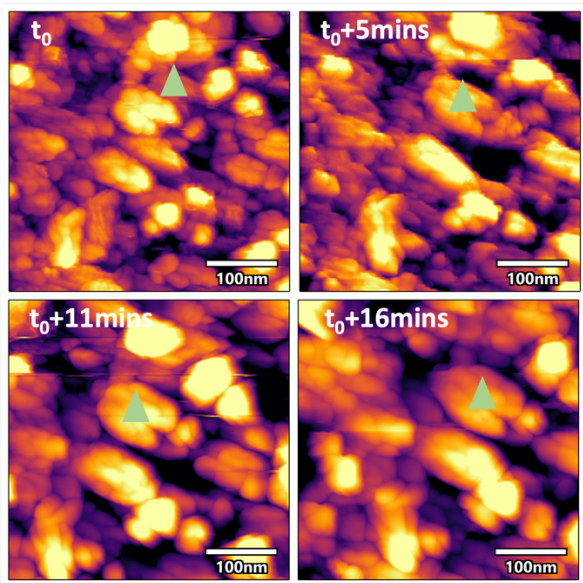

**Figure S10.** Selected snapshots of *in-situ* AFM of Zn metal in water show that the Zn surface reorganized from the granular structure to a new structure with larger grain sizes ($t_0$ is 30 minutes after soaking in water).

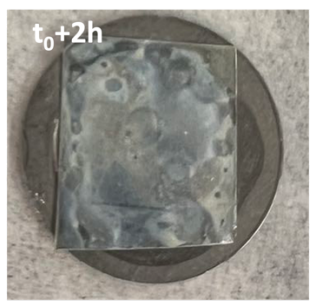

**Figure S11. The picture of Zn metal in water after two hours.** Zn metal shows greyish film and pits on the surface.

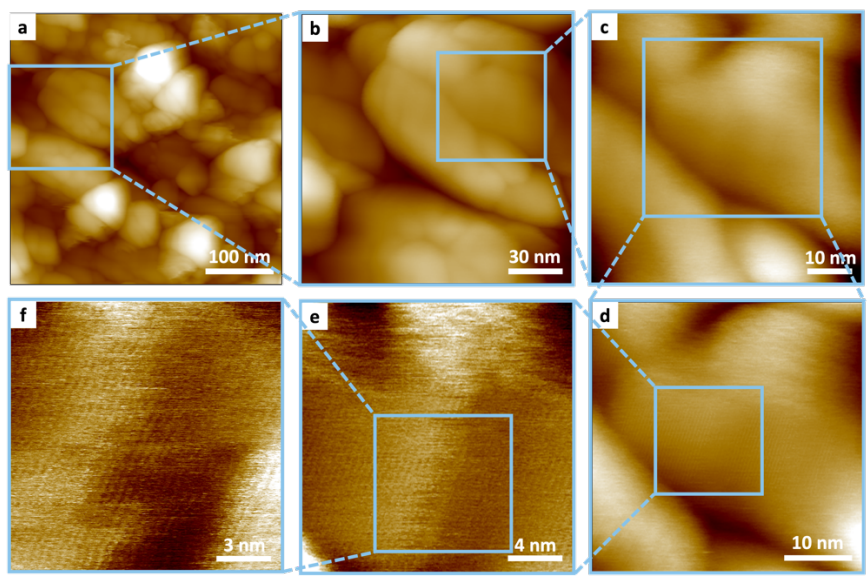

**Figure S12**. The zoomed-in height image of the new structure formed by soaking Zn metals in water for 60 mins.

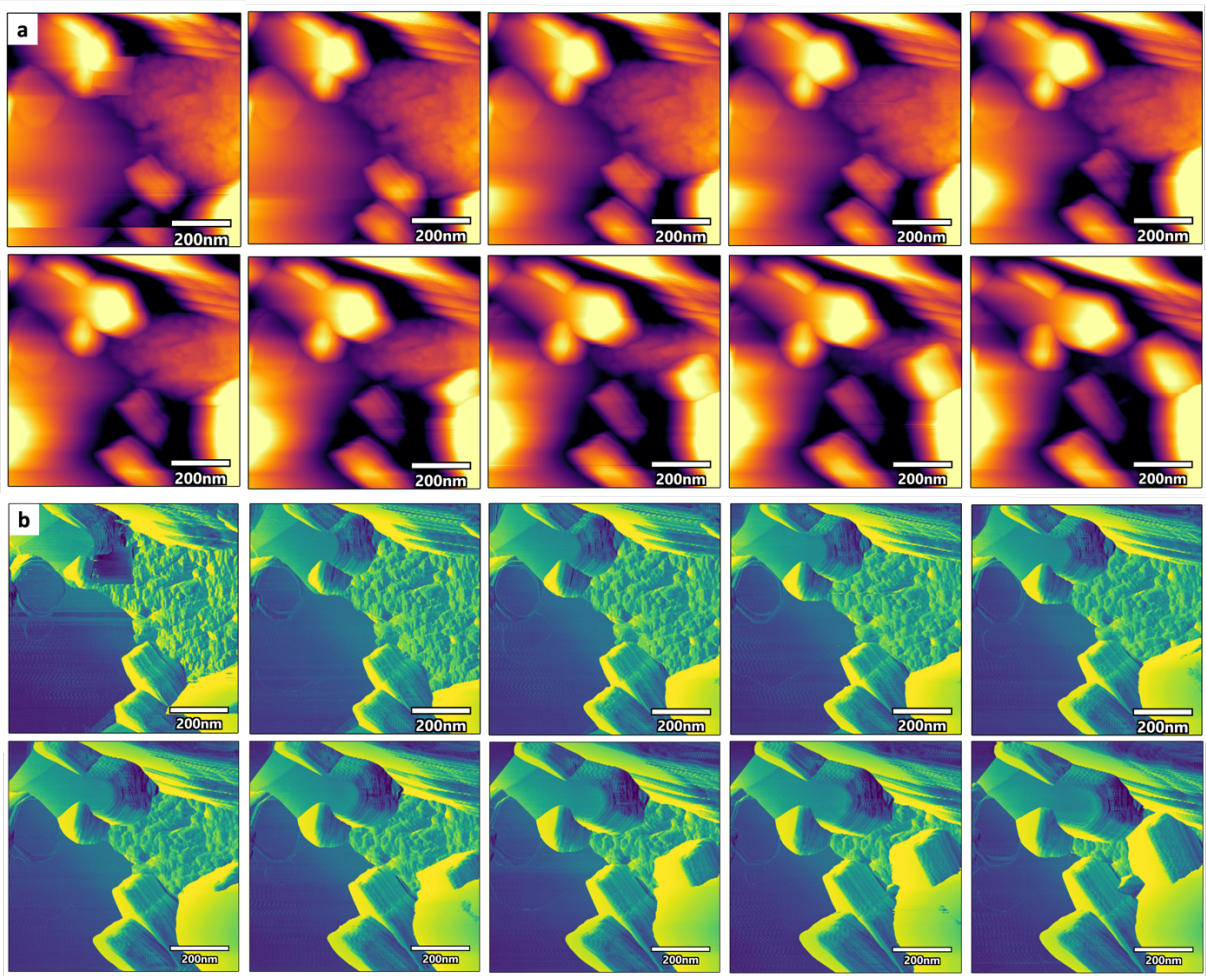

**Figure S13**. Selected snapshots of *in-situ* height (a) and phase (b) images of Zn metal in 0.1 M ZnSO₄ show the Zn surface reorganized into the hexagonal layered structure. (images with intervals 65 s).

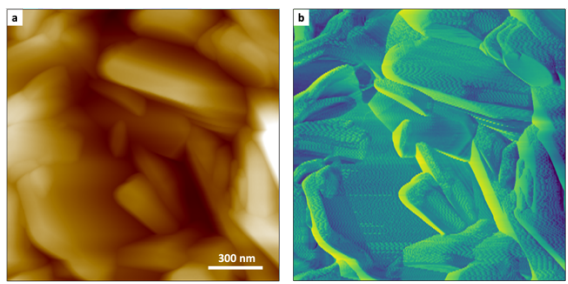

**Figure S14**. The large scale of height (a) and phase (b) images of Zn metal soaking in 0.1 M ZnSO₄ after 2 hours shows the Zn surface fully reorganized into the hexagonal layered structure with big roughness. (height range: 1000 nm)

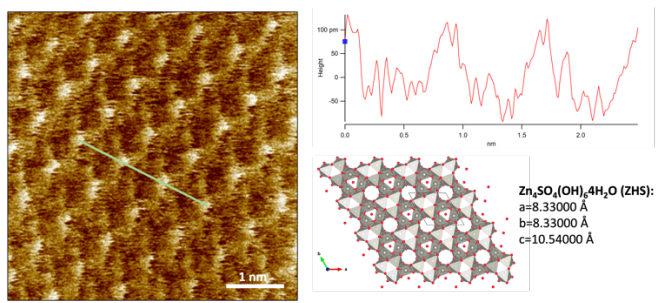

**Figure S15**. The high-resolution AFM image with the height line profile shows the lattice parameter of the observed structure of Zn soaking in 0.1 M ZnSO₄ (a = c = 0.833 nm), which is comparable to the that of ZHS found in the database.

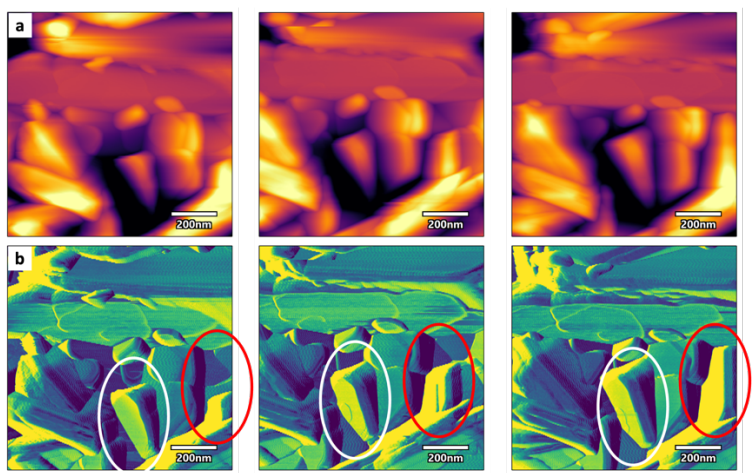

**Figure S16**. Selected snapshots of *in-situ* height (a) and phase (b) images of Zn metal in 0.1 M ZnSO₄ with 0.1 wt% PEO show the reorganization of the Zn surface into the hexagonal layered structure is not hindered by PEO additives. White and red circles highlight regions showing significant surface reorganization. (left: $t_0$; middle: $t_0$ +10 mins; right: $t_0$+20 mins.)

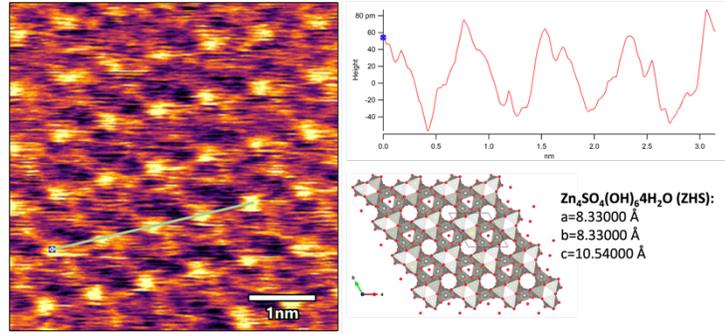

**Figure S17**. The high-resolution AFM image with the height line profile shows the lattice parameters of the observed structure of Zn soaking in 0.1 M ZnSO4 with 0.1 wt% PEO (a = c =0.8 nm), which matches with that of ZHS found in the database.

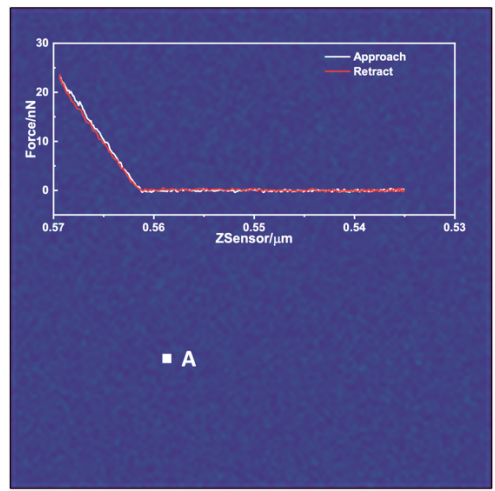

**Figure S18**. The adhesion force map of Zn metal in 0.1 M ZnSO$_4$ solution and single point force curve of point A analyzed from the adhesion force map was inserted.

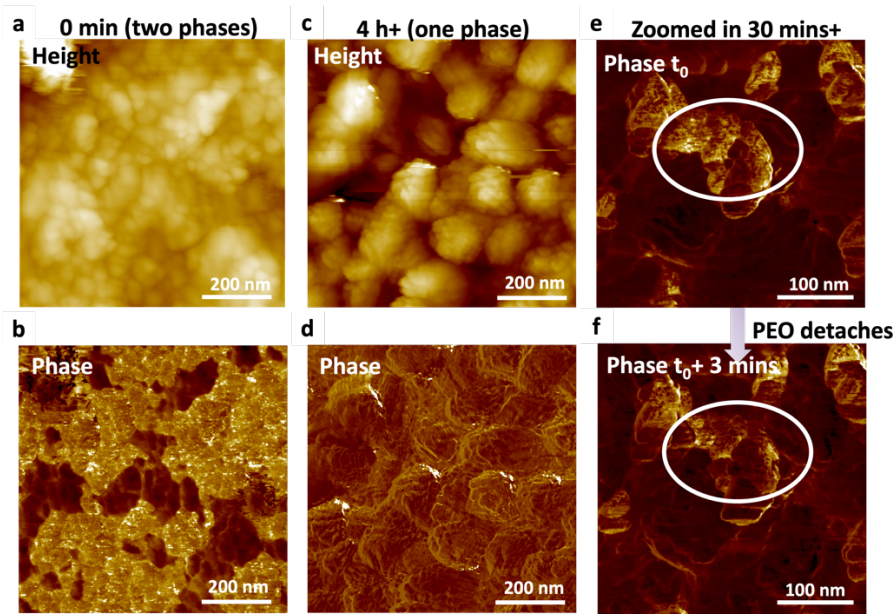

**Figure S19. AFM results of Zn foil in 0.1 wt% PEO solution imply that PEO detaches from Zn due to surface reorganization of Zn.** (a, b) The height (a) and correlated phase (b) images at the early stage of Zn foil soaking in PEO solution. (c, d) The height (c) and correlated phase (d) images were captured 4 hours later than (a, b). (e, f) The zoomed-in phase images captured 30 minutes after (a, b). (f) was captured 3 minutes after (e). White circles highlight the regions with both the phase contrast change and the morphology change.

We adopted the protocol to correlate phase and height images to visualize the PEO polymer adsorption and distribution on Zn surfaces. Our results show that PEO will detach from the Zn surface due to surface reorganization of Zn. In the initial stage of Zn foil soaking in 0.1 wt% PEO solution, two phases coexist on the rough Zn foil surface (**Fig. S19a, S19b**). Four hours later, the morphology evolved to bundles of platelet structure with a bigger grain size. The phase changed from two phases coexisting to only one phase (**Fig. S19c, S19d**). The phase images with higher resolution were conducted to understand the underlying mechanisms. From the region highlighted by white circles, we can see the lighter region diminished when the surface was reorganized, which means PEO detaching from the Zn surface due to Zn surface reorganization (**Fig. S19e, S19f**).

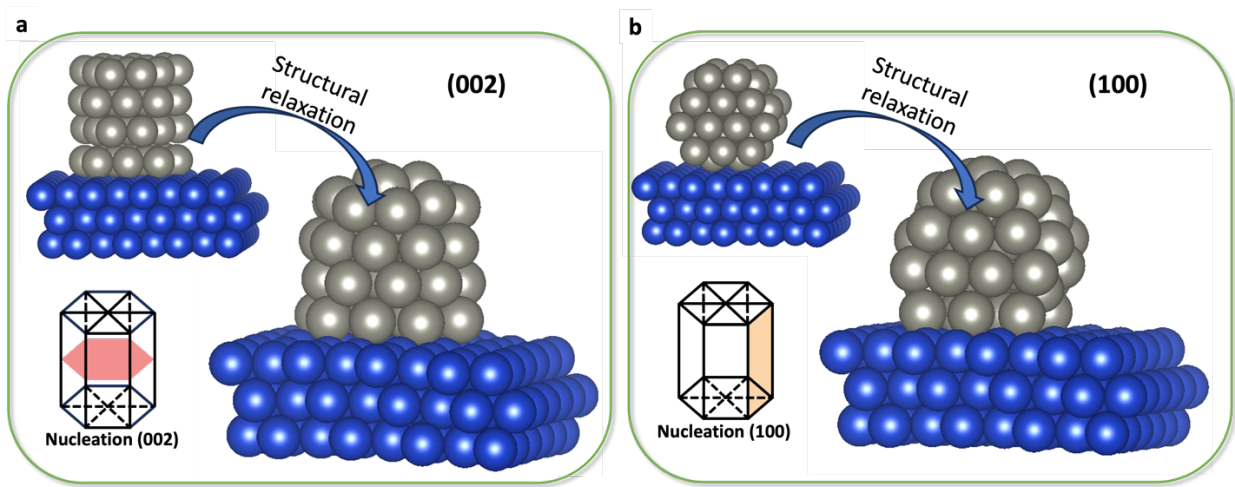

**Figure S20**. The DFT results show that the nucleation of the Zn (002) plane is less preferable than the nucleation of the Zn (100) on Cu in vacuum. The energy per surface atom is 1.2 times lower for (100) orientation than that for (002) plate. These structures are shown in ball-and-stick form, with Cu and Zn atoms represented in blue and gray, respectively.

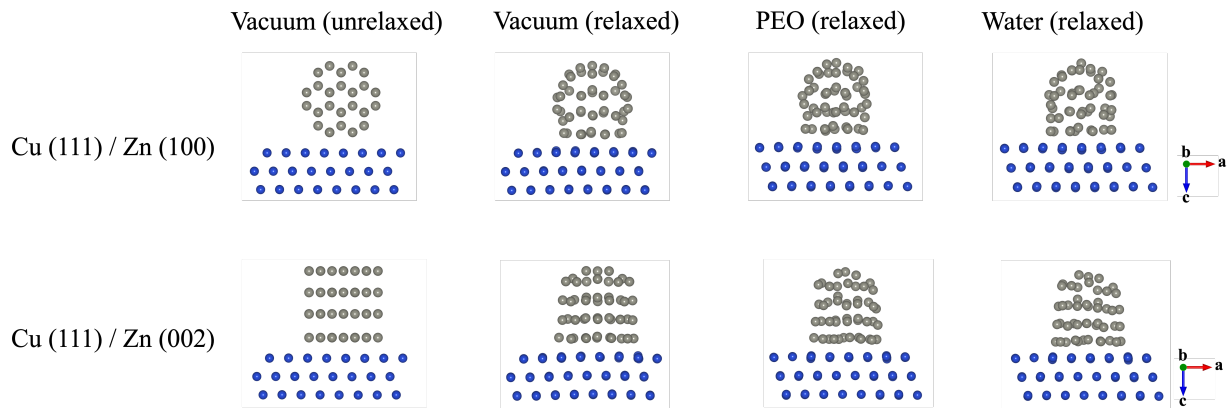

**Figure S21.** Structures of unrelaxed and relaxed Cu (111) / Zn (100) and Cu (111) / Zn (002) interfaces in vacuum, PEO, and water (solvent molecules removed). These structures are shown in ball-and-stick form, with Cu and Zn atoms represented in blue and gray, respectively.

**Table S1.** Average interatomic distances (Å) between Zn atoms in the first layer adjacent to the Cu slab at the Cu (111) / Zn (100) and Cu (111) / Zn (002) interfaces in vacuum, PEO, and water.

|  | Vacuum (unrelaxed) | Vacuum (relaxed) | PEO (relaxed) | Water (relaxed) |
| --- | --- | --- | --- | --- |
| Cu (111) / Zn (100) | 2.614 | 3.065 | 2.927 | 2.956 |
| Cu (111) / Zn (002) | 2.614 | 2.948 | 2.677 | 2.813 |

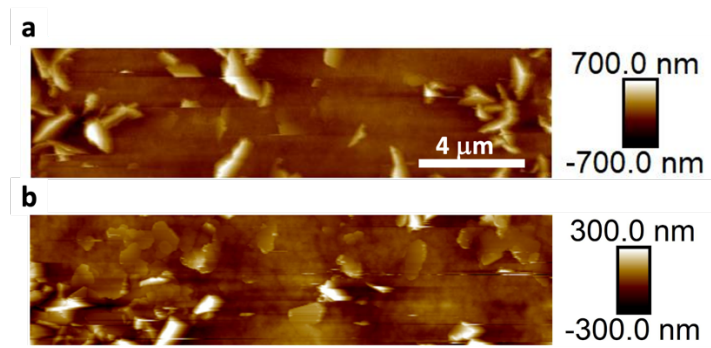

**Figure S22.** *In-situ* **EC-AFM results of Zn electrodeposited on Cu in 0.01 M ZnSO$_4$ with and without PEO additives.** (a) 0.01 M ZnSO$_4$, (b) 0.01 M ZnSO$_4$ with 0.1 wt% PEO. The final deposited plates in the presence of PEO show more parallel-oriented plates compared to those formed without PEO.

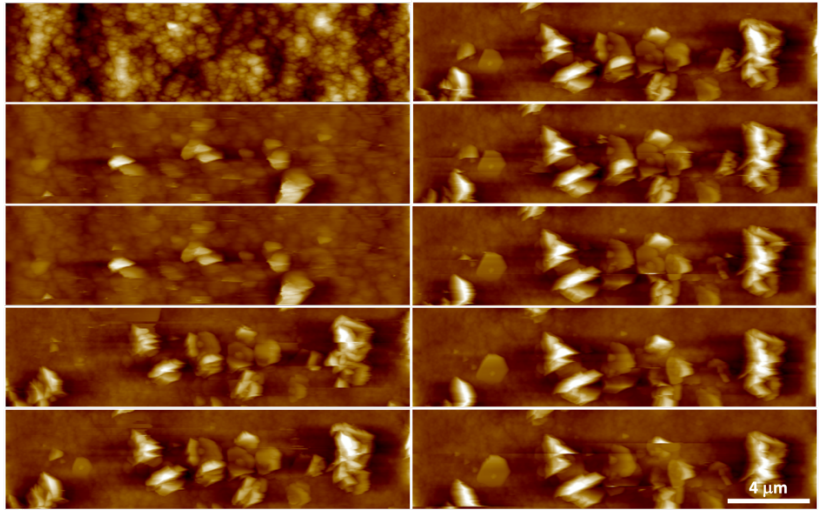

**Figure S23.** *In-situ* AFM images of Zn electrodeposition on Cu in 0.1 M ZnSO$_4$ after nitrogen bubbling to remove dissolved oxygen.

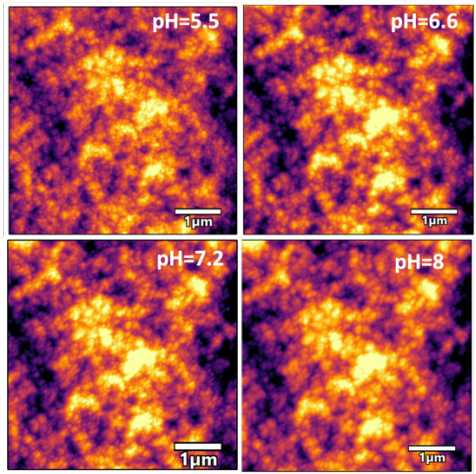

**Figure S24**. *In-situ* flowing AFM images with increasing pH show no plates deposited on Cu surfaces.

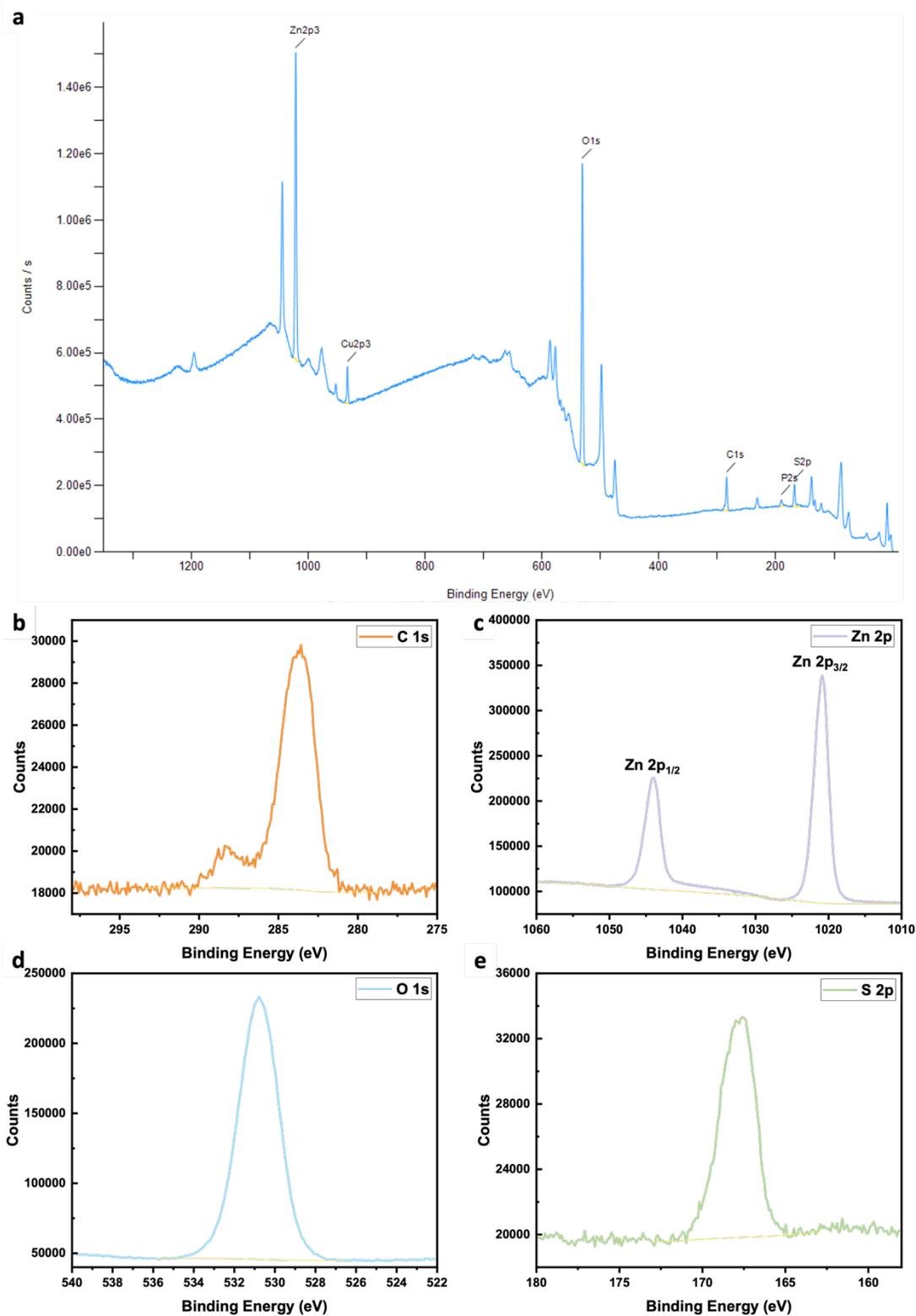

**Figure S25.** Ex-situ XPS of Zn deposited on Cu in 0.1 M ZnSO$_4$ solution. (a) A broad scan of the sample. High-resolution XPS spectra of C 1s (b), Zn 2p (c), O1s (d), and S 2p (e) orbitals.

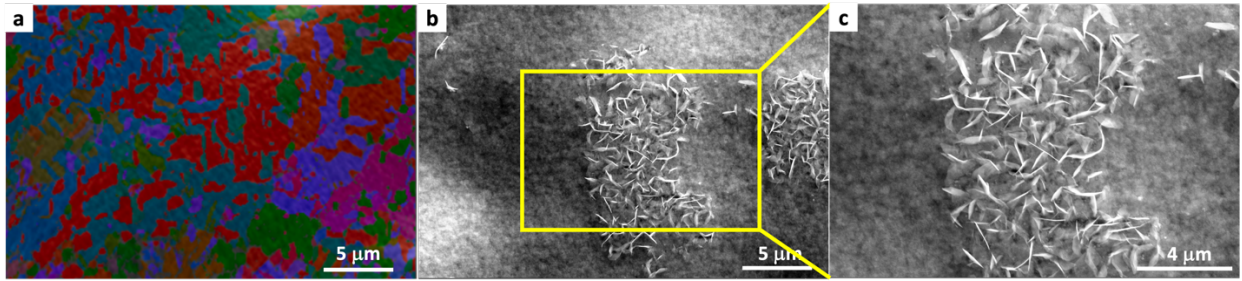

**Figure S26**. EBSD images of Cu substrate before Zn electrodeposition (a) and SEM images (b, c) of Zn on Cu after 30 seconds of electrodeposition.

## Section 4. Caption for Video S1 and Video S2

**Video S1:** *In-situ* AFM results of Zn metal in water show the rapid surface reorganization of Zn when soaked in water (65 s / frame)

**Video S2:** *In-situ* AFM results of Zn metal in 0.1 M $ZnSO_4$ show the rapid surface reorganization of Zn when soaked in $ZnSO_4$ solution (65 s / frame)

## Section 5. References

[1] J. S. Keist, C. A. Orme, P. K. Wright, J. W. Evans, *Electrochimica Acta* **2015**, *152*, 161–171.
[2] A. D. Becke, *The Journal of Chemical Physics* **2014**, *140*, 18A301.
[3] E. Bylaska, K. Tsemekhman, N. Govind, M. Valiev, in *Computational Methods for Large Systems*, **2011**, pp. 77–116.
[4] M. Valiev, E. J. Bylaska, N. Govind, K. Kowalski, T. P. Straatsma, H. J. J. Van Dam, D. Wang, J. Nieplocha, E. Apra, T. L. Windus, W. A. de Jong, *Computer Physics Communications* **2010**, *181*, 1477–1489.
[5] E. Aprà, E. J. Bylaska, W. A. de Jong, N. Govind, K. Kowalski, T. P. Straatsma, M. Valiev, H. J. J. van Dam, Y. Alexeev, J. Anchell, V. Anisimov, F. W. Aquino, R. Atta-Fynn, J. Autschbach, N. P. Bauman, J. C. Becca, D. E. Bernholdt, K. Bhaskaran-Nair, S. Bogatko, P. Borowski, J. Boschen, J. Brabec, A. Bruner, E. Cauët, Y. Chen, G. N. Chuev, C. J. Cramer, J. Daily, M. J. O. Deegan, T. H. Dunning Jr., M. Dupuis, K. G. Dyall, G. I. Fann, S. A. Fischer, A. Fonari, H. Früchtl, L. Gagliardi, J. Garza, N. Gawande, S. Ghosh, K. Glaesemann, A. W. Götz, J. Hammond, V. Helms, E. D. Hermes, K. Hirao, S. Hirata, M. Jacquelin, L. Jensen, B. G. Johnson, H. Jónsson, R. A. Kendall, M. Klemm, R. Kobayashi, V. Konkov, S. Krishnamoorthy, M. Krishnan, Z. Lin, R. D. Lins, R. J. Littlefield, A. J. Logsdail, K. Lopata, W. Ma, A. V. Marenich, J. Martin del Campo, D. Mejia-Rodriguez, J. E. Moore, J. M. Mullin, T. Nakajima, D. R. Nascimento, J. A. Nichols, P. J. Nichols, J. Nieplocha, A. Otero-de-la-Roza, B. Palmer, A. Panyala, T. Pirojsirikul, B. Peng, R. Peverati, J. Pittner, L. Pollack, R. M. Richard, P. Sadayappan, G. C. Schatz, W. A. Shelton, D. W. Silverstein, D. M. A. Smith, T. A. Soares, D. Song, M. Swart, H. L. Taylor, G. S. Thomas, V. Tipparaju, D. G. Truhlar, K. Tsemekhman, T. Van Voorhis, Á. Vázquez-Mayagoitia, P. Verma, O. Villa, A. Vishnu, K. D. Vogiatzis, D. Wang, J. H. Weare, M. J. Williamson, T. L. Windus, K. Woliński, A. T. Wong, Q. Wu, C. Yang, Q. Yu, M. Zacharias, Z. Zhang, Y. Zhao, R. J. Harrison, *The Journal of Chemical Physics* **2020**, *152*, 184102.
[6] W. Kohn, L. J. Sham, *Phys. Rev.* **1965**, *140*, A1133–A1138.

[7] D. R. Hamann, *Phys. Rev. B* **1989**, *40*, 2980–2987.
[8] N. Troullier, J. L. Martins, *Phys. Rev. B* **1991**, *43*, 1993–2006.
[9] L. Kleinman, D. M. Bylander, *Phys. Rev. Lett.* **1982**, *48*, 1425–1428.
[10] D. Meng, G. Lin, M. L. Sushko, *MRS Online Proceedings Library* **2012**, *1470*, 1–5.
[11] D. Meng, B. Zheng, G. Lin, M. L. Sushko, *Communications in Computational Physics* **2014**, *16*, 1298–1322.
[12] J. Wu, Z. Li, *Annual Review of Physical Chemistry* **2007**, *58*, 85–112.
[13] R. Roth, *Journal of Physics: Condensed Matter* **2010**, *22*, 063102.
[14] Y.-X. Yu, J. Wu, *The Journal of Chemical Physics* **2002**, *117*, 10156–10164.
[15] J. S. Høye, L. Blum, *Molecular Physics* **1978**, *35*, 299–300.
[16] L. Blum, *Molecular Physics* **1975**, *30*, 1529–1535.
[17] N. Grønbech-Jensen, R. J. Mashl, R. F. Bruinsma, W. M. Gelbart, *Phys. Rev. Lett.* **1997**, *78*, 2477–2480.
[18] I. Rouzina, V. A. Bloomfield, *J. Phys. Chem.* **1996**, *100*, 9977–9989.
[19] J. J. Arenzon, J. F. Stilck, Y. Levin, *The European Physical Journal B - Condensed Matter and Complex Systems* **1999**, *12*, 79–82.
[20] Y. Levin, J. J. Arenzon, J. F. Stilck, *Phys. Rev. Lett.* **1999**, *83*, 2680–2680.
[21] Y. W. Kim, J. Yi, P. A. Pincus, *Phys. Rev. Lett.* **2008**, *101*, 208305.
[22] G. S. Manning, *The European Physical Journal E* **2011**, *34*, 132.
[23] J. C. Butler, T. Angelini, J. X. Tang, G. C. L. Wong, *Phys. Rev. Lett.* **2003**, *91*, 028301.
[24] Q. Liao, A. V. Dobrynin, M. Rubinstein, *Macromolecules* **2003**, *36*, 3386–3398.
[25] M. Sedlák, E. J. Amis, *The Journal of Chemical Physics* **1992**, *96*, 826–834.
[26] Y. Jin, K. S. Han, Y. Shao, M. L. Sushko, J. Xiao, H. Pan, J. Liu, *Advanced Functional Materials* **2020**, *30*, 2003932.
[27] J. P. Healy, D. Pletcher, M. Goodenough, *Journal of Electroanalytical Chemistry* **1992**, *338*, 155–165.
[28] C. Newey, G. Weaver, *Materials Principles and Practice: Electronic Materials Manufacturing with Materials Structural Materials*, Elsevier, **2013**.
[29] A. Jain, S. P. Ong, G. Hautier, W. Chen, W. D. Richards, S. Dacek, S. Cholia, D. Gunter, D. Skinner, G. Ceder, *APL materials* **2013**, *1*.